\documentclass[11pt]{article}

\usepackage[final]{acl}

\usepackage{times}
\usepackage{latexsym}

\usepackage[T1]{fontenc}
\usepackage[utf8]{inputenc}
\usepackage{amsmath} 
\usepackage{amssymb}

\usepackage{multirow} 
\usepackage{microtype}
\usepackage{tabularx}
\usepackage{booktabs}
\usepackage{inconsolata}

\usepackage{graphicx}
\usepackage{float}
\usepackage{bm} 
\usepackage[utf8]{inputenc}
\usepackage{amsmath} 

\title{FinRankGRPO: Optimizing LLMs for Listwise Financial Asset Ranking via Group Relative Policy Optimization}
\author{
  \textbf{Ningyuan Deng\textsuperscript{1}},
  \textbf{Jinyuan Wang\textsuperscript{1}},
  \textbf{Qi Li\textsuperscript{2}},
  \textbf{Yuxin Wei\textsuperscript{2}},
  \textbf{Jia Zhang\textsuperscript{2}},
  \textbf{Yi Yang\textsuperscript{1}}
\\
  \textsuperscript{1}The Hong Kong University of Science and Technology, \\
  \textsuperscript{2}Ping An Property \& Casualty Insurance Company of China, Ltd
\\[1em]
  \begin{tabular}{c}
    \href{mailto:ndengad@connect.ust.hk}{ndengad@connect.ust.hk},
    \href{mailto:jwangiy@connect.ust.hk}{jwangiy@connect.ust.hk},
    \href{mailto:li.qi@graduate.utm.my}{li.qi@graduate.utm.my} \\
    \href{mailto:weiyuxin033@pingan.com.cn}{weiyuxin033@pingan.com.cn},
    \href{mailto:zhangjia348@pingan.com.cn}{zhangjia348@pingan.com.cn},
    \href{mailto:imyiyang@ust.hk}{imyiyang@ust.hk}
  \end{tabular}
}

\begin{document}
\maketitle
\begin{abstract}

While large language models (LLMs) excel at understanding unstructured financial contexts, their direct use in portfolio optimization is limited by a mismatch between next-token prediction and the listwise ranking objectives required for asset allocation. They also struggle to produce reliable numerical forecasts, leading to instability and arithmetic hallucinations. To bridge this gap, we propose FinRankGRPO, a framework that shifts LLM-based portfolio construction from direct numerical prediction to listwise asset ranking. Our method employs a two-stage training process consisting of supervised fine-tuning on chain-of-thought reasoning data, followed by Financial Asset Ranking via Group Relative Policy Optimization with a Spearman-based reward that encourages agreement between generated rankings and ground-truth market orderings. The second stage provides graded listwise feedback rather than sparse exact-match rewards. Experimental results show that FinRankGRPO achieves the highest Spearman correlation coefficient and the highest mean Sharpe ratio among the evaluated traditional quantitative baselines and commercial LLMs, reaching 0.018 and 0.689, respectively, on a 15-ETF universe. Our code and data are publicly available at~\url{https://anonymous.4open.science/r/arr_10_finrankgrpo-7346/}.
\end{abstract}

\section{Introduction}

\begin{figure}[htbp]
    \centering
    \includegraphics[width=\linewidth]{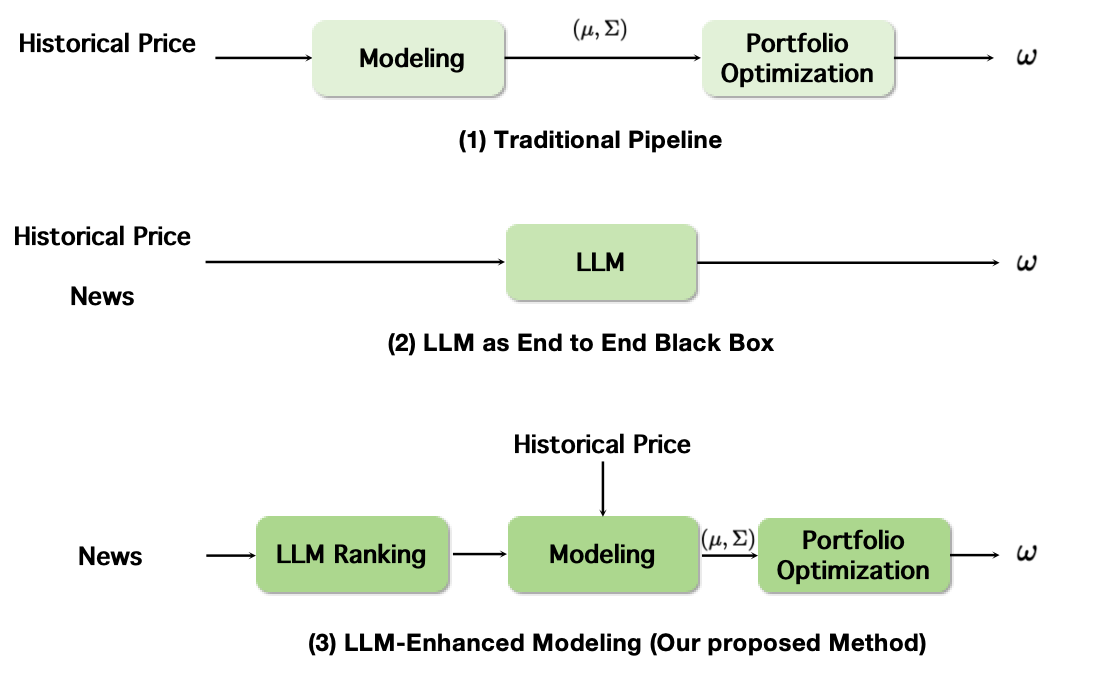}
    % \caption{Portfolio construction pipeline. The data modeling block estimates the expected return vector $\mu$ and covariance matrix $\Sigma$ of the future return distribution, which serve as inputs to the portfolio optimization block.}
    % \caption{Comparison of portfolio management approaches. $\mu$ denotes the expected return vector and $\Sigma$ denotes the asset return covariance matrix. (1) The traditional two stage pipeline separates return modeling and portfolio optimization. (2)LLM as End-to-end Black box leverages LLMs directly to map inputs, including market news and historical prices, to portfolio weights. (3) Our proposed method, instead of directly using an LLM to produce asset weights or numerical estimates of expected returns, generates a listwise ranking of assets, which is subsequently incorporated into the portfolio optimization stage.}
    \caption{Comparison of portfolio management approaches. $\mu$ and $\Sigma$ denote the expected returns and the covariance matrix, respectively. (1) Traditional pipelines separate return modeling and portfolio optimization. (2) LLM as End-to-end Black Box. (3)  Our method generates listwise asset rankings and incorporates them into portfolio optimization.}
    \label{fig:portfolio_optimization_pipeline}
\end{figure}

Portfolio management is a cornerstone of financial investment, centered on allocating capital across different assets (equities, bonds, commodities, and alternatives) to maximize returns while minimizing risk~\cite{Markowitz1971PortfolioS, Fabozzi2007RobustPO, Kirtac2025LeveragingLS}. Classical approaches such as Mean-Variance Optimization (MVO)~\cite{markowitz1952modern} provide a standard backbone for portfolio construction given estimates of return and covariance. Later, the Black-Litterman and Entropy Pooling (EP) frameworks extend this paradigm by allowing investors to incorporate subjective views, such as relative outperformance expectations, into the portfolio construction process ~\cite{black1992global, idzorek2007step, Meucci2008FullyFV}, as illustrated as Path 1 in Figure~\ref{fig:portfolio_optimization_pipeline}. However, these frameworks assume structured numerical views (e.g., ``asset A is expected to outperform asset B by 5\%'') and offer limited guidance on deriving such views, or relative asset rankings, from unstructured text such as market news ~\cite{Mahdavi2025IntegratingLL, Lee2025LLMEnhancedBP}.

In recent years, Large Language Models (LLMs) have emerged as a promising solution, as shown in Path 2 of Figure~\ref{fig:portfolio_optimization_pipeline}, by integrating textual and numerical data for financial analysis. FinSrag~\cite{Xiao2025RetrievalaugmentedLL} uses retrieval-augmented generation (RAG) for financial time series forecasting. LLM-BLM~\cite{Lee2025LLMEnhancedBP} integrates LLM-generated prediction returns and confidence into the Black-Litterman model. Agent-based architectures, like MarketSenseAI~\cite{Fatouros2025MarketSenseAI2E}, employ specialized AI agents to analyze multi-source financial data and produce stock selection signals. Despite their advances, existing LLM-augmented approaches still face two critical limitations.

% Firstly, there is a fundamental mismatch between the objectives of standard LLM training and the nature of financial investment. Most existing LLM-based approaches are not designed to optimize ranking relationships. Recent supervised fine‑tuning (SFT) efforts attempt to adapt LLMs to direct stock price~\cite{Guo2024FineTuningLL} or return forecasting~\cite{Chiu2024FinancespecificLL}; however, these methods are still based on next‑token prediction losses, which do not explicitly capture the list-wise ranking structures essential for portfolio construction. Similarly, agent-based frameworks still rely on generic training paradigms that do not optimize for ordinal accuracy. For example, TradingAgents~\cite{Xiao2024TradingAgentsML} employs specialized LLM agents with varying risk profiles, and FinCon~\cite{Yu2024FinConAS} adopts a manager–analyst communication hierarchy to mimic real-world investment firms. Reinforcement learning methods also exhibit this limitation. Group Relative Policy Optimization (GRPO)~\cite{Shao2024DeepSeekMathPT} is initially designed for binary verification in mathematical or coding tasks and lacks a mechanism to learn rankings. 
% Although SAPPO~\cite{Kirtac2025LeveragingLS}enhances portfolio optimization by integrating reinforcement learning with sentiment analysis, its reward mechanism lacks ranking-specific objectives. Consequently, none of the aforementioned approaches directly encourages models to produce economically meaningful asset rankings, which is a core requirement for effective and interpretable portfolio construction.

Firstly, there is a fundamental mismatch between standard LLM training objectives and financial investment. Most LLM-based approaches are not designed to optimize ranking relationships. Recent supervised fine-tuning (SFT) methods adapt LLMs to stock price~\cite{Guo2024FineTuningLL} or return forecasting~\cite{Chiu2024FinancespecificLL}, but still rely on next token prediction losses that fail to capture the listwise ranking structures required for portfolio construction. Agent-based frameworks, such as TradingAgents~\cite{Xiao2024TradingAgentsML} and FinCon~\cite{Yu2024FinConAS}, also rely on generic training paradigms rather than ordinal objectives. Reinforcement learning methods share this limitation. GRPO~\cite{Shao2024DeepSeekMathPT} is designed for binary verification rather than ranking, while SAPPO~\cite{Kirtac2025LeveragingLS} enhances portfolio optimization by integrating reinforcement learning with sentiment analysis. Its reward mechanism lacks ranking-specific objectives for portfolio management. Consequently, existing methods do not directly encourage economically meaningful asset rankings, which are essential for effective and interpretable portfolio construction.

Secondly, LLMs have limitations in reliable numerical forecasting, which weakens their direct use in quantitative portfolio management. Recent studies~\cite{Lee2025YourAN} show that this approach is susceptible to arithmetic hallucinations and investment biases~\cite{Lee2025YourAN}. Even advanced models such as OpenAI o3-deep-research achieve only approximately 40$\%$ accuracy, suggesting that internal reasoning is still insufficient for financial numeracy forecasting~\cite{Li2026FinDeepForecastAL}. Instead of relying on brittle point estimates, ordinal ranking avoids generating calibrated return magnitudes, which are noisy and difficult to obtain, and instead uses the LLM's comparative reasoning to produce relative portfolio views directly.

% This raises a natural question: Can we design a ranking driven reinforcement learning framework that adapts LLMs to the asset ranking task for portfolio construction?

This raises a natural question: Can we design a ranking driven reinforcement learning framework that adapts LLMs to asset ranking for portfolio construction?

% To answer this question, we move from directly using LLMs for numerical predictions to leveraging them for robust asset ranking. Specifically, we introduce the Financial Asset Ranking via Group Relative Policy Optimization (FinRankGRPO) on Path 3 of Figure~\ref{fig:portfolio_optimization_pipeline}, a novel reinforcement learning framework that optimizes LLM using a ranking-based reward. This reward is based on Spearman's correlation between the model's predicted asset order and the true performance ordering. By converting financial reasoning into a listwise ranking problem, FinRankGRPO directly aligns the LLM with the ordinal structure of investment decisions. These learned rankings are then integrated into an Entropy Pooling (EP) framework~\cite{meucci2016dynamic, meucci2010fully} as relative preference constraints over assets. %This design allows unstructured textual information to influence portfolio construction without requiring direct numerical forecasts of returns. 

To answer this question, we move from numerical prediction to robust asset ranking. Specifically, We introduce FinRankGRPO, shown as Path 3 in Figure ~\ref{fig:portfolio_optimization_pipeline}, which is a reinforcement-learning framework that optimizes LLMs using a ranking-based reward. The reward is based on Spearman's correlation between the predicted asset order and the ground-truth ordering. By converting financial reasoning into a listwise ranking problem, FinRankGRPO aligns LLMs with the ordinal structure of investment decisions. The learned rankings are then integrated into an EP framework~\cite{meucci2016dynamic, meucci2010fully} as relative preference constraints over assets.

% we introduce Financial Asset Ranking via Group Relative Policy Optimization (FinRankGRPO), shown as Path 3 in Figure, a reinforcement learning framework that optimizes LLMs with a ranking based reward

% \item \textbf{The FinRankGRPO Framework:} We propose a novel alignment paradigm that reformulates financial text understanding as a listwise ranking task and use FinRankGRPO with a Spearman‑based reward to enforce structural and ordinal accuracy.

% \item \textbf{Integration LLM with EP} We develop an integration of LLM-derived rankings into an Entropy Pooling framework to ensure that textual insights are grounded by mathematical rigor without numerical instability. 

% \item \textbf{Empirical Superiority:} Through extensive experiments, our results show that FinRankGRPO significantly outperforms traditional quantitative baselines, improving both ranking metrics and financial performance. 

% Our contributions are summarized as follows. First, the FinRankGRPO Framework: we propose a novel alignment paradigm that reformulates financial text understanding as a listwise ranking task and uses a Spearman-based reward to enforce structural and ordinal accuracy. Second, {Integrating LLMs with EP}, we integrate LLM-derived rankings into an Entropy Pooling framework, grounding textual insights in mathematical portfolio optimization without numerical instability. Third, {Empirical Superiority}, extensive experiments show that FinRankGRPO significantly outperforms traditional quantitative and LLM baselines, improving both ranking metrics and financial performance.

Our contributions are summarized as follows. First, we propose FinRankGRPO, a ranking-based alignment framework that reformulates financial text understanding as a listwise asset ranking task and optimizes LLMs with a Spearman-based reward for structural and ordinal accuracy. Second, we integrate LLM-derived rankings into an EP framework, enabling textual signals to inform mathematically grounded portfolio optimization without requiring direct numerical forecasts. Third, we conduct experiments on a 15-ETF universe, where FinRankGRPO achieves the highest mean Spearman correlation, annualized return, and Sharpe ratio among the evaluated methods.

% Third, we
% conduct extensive experiments showing that FinRankGRPO outperforms traditional quantitative
% and LLM baselines in both ranking quality and
% portfolio performance.

%Third, we evaluate this design in a restricted ETF and stock setting and analyze its ranking quality, portfolio utility, and transfer boundary on an individual-stock universe.

\section{Related Work}

\subsection{LLMs in Portfolio Management}

Modern Portfolio Theory frames portfolio construction as an optimization between expected return and variance risk~\cite{markowitz1952modern, Fabozzi2007RobustPO}, but it is sensitive to noisy return estimates~\cite{idzorek2007step, black1992global} and can overemphasize assets with estimation errors~\cite{michaud1989markowitz}. The Black-Litterman model mitigates this issue by incorporating subjective views into posterior return and covariance estimates~\cite{black1992global, idzorek2007step}, while EP further generalizes view integration and supports ordinal ranking constraints~\cite{meucci2010fully, mercurio2020entropy}.

LLMs have recently expanded financial text understanding beyond conventional forecasting. General models such as ChatGPT~\cite{Dowling2023ChatGPTF}, domain models such as BloombergGPT~\cite{Wu2023BloombergGPTAL} and FinGPT~\cite{Yang2023FinGPTOF}, and reasoning systems such as Fin-R1~\cite{Liu2025FinR1AL} and DianJin-R1~\cite{Zhu2025DianJinR1EA} show the value of domain adaptation, instruction tuning, and reinforcement learning for financial reasoning. Agent frameworks, including FinAgent~\cite{zhang2024multimodal} and FinCon~\cite{yu2024fincon}, further broaden these applications.

However, financial LLMs are still mostly used for stock picking~\cite{xiao2024tradingagents}, buy/sell/hold decisions~\cite{zhang2024multimodal}, or numerical views for Black-Litterman optimization~\cite{Lee2025LLMEnhancedBP}. FinRankGRPO instead uses LLMs to produce predictive ranking views, decoupling textual reasoning from direct numerical return or weight generation. This design reduces numerical inconsistency while preserving interpretable Chain-of-Thought (CoT) rationales for allocation decisions.

\subsection{LLM-Based Ranking}
LLM-based ranking has been explored in information retrieval~\cite{zhang2024ai, khramtsova2024leveraging, Sun2023IsCG}, recommendation~\cite{gao2025llm4rerank, sharma2024optimizing, Zhu2025RankGRPOTL}, and question answering~\cite{dehghankar2025rank, Pradeep2023RankVicunaZL}. Early methods encode items as discrete tokens and cast ranking as sequence generation~\cite{Hua2023HowTI, zhu2024collaborative}, while later prompt-based methods generate recommendation lists in natural language~\cite{Zhu2025CollaborativeRF}. Systems such as Chat-Rec~\cite{Gao2023ChatRECTI} and InstructRec~\cite{Zhang2023RecommendationAI} show that instruction tuned LLMs can follow complex ranking criteria, but their outputs may remain unstable without explicit ranking optimization. Recent work, such as RankGRPO~\cite{Zhu2025RankGRPOTL}, addresses this limitation by aligning LLM outputs with ranking specific rewards through Group Relative Policy Optimization.

Applying ranking optimized LLMs to financial asset allocation remains challenging because market signals are noisy and rankings must align with realized dynamics rather than static preferences. FinRankGRPO addresses this gap with a two-stage, reward driven framework for listwise financial asset ranking, grounding LLM reasoning in quantitative market performance through structured preference learning.

\section{Method}

\subsection{Problem Formulation}

We formulate portfolio management as a two-stage process: (1) extracting predictive asset rankings from unstructured text such as news, and (2) mapping these rankings to valid portfolio weights. Given news up to the weekly rebalancing time \(t\), the model generates a full permutation of the \(N\) assets. The target permutation is obtained by sorting assets by their realized returns over the next one-week holding period \((t,t+1]\).

\textbf{Ranking Generation.}
Formally, given a set of \(N\) financial assets
\(\mathcal{A}=\{a_1,\ldots,a_N\}\), financial news
\(X_{\text{news}}(t)\), and the corresponding trading date
\(X_{\text{date}}(t)\), the LLM generates a textual response
\(y_{t+1}\):
\begin{equation}
y_{t+1}
=
\operatorname{LLM}_{\theta}
\left(
\mathcal{T}
\left(
X_{\text{date}}(t),
X_{\text{news}}(t),
\mathcal{A}
\right)
\right),
\end{equation}
% where \(\mathcal{T}(\cdot)\) denotes the prompt template used to elicit a complete asset ranking, as shown in Appendix~\ref{appx:prompt}, and \(\theta\) denotes the parameters of the LLM. A deterministic parser extracts the predicted permutation \(\hat{\pi}_{t+1}\) from \(y_{t+1}\).

where \(\mathcal{T}\) is the ranking prompt template (Appendix~\ref{appx:prompt}) and \(\theta\) the LLM parameters; a deterministic parser extracts \(\hat{\pi}_{t+1}\) from \(y_{t+1}\).
\textbf{Portfolio Construction.} 
% The generated ranking $y_{t+1}$ is qualitative. Our goal is to map this discrete ranking into continuous portfolio weights $\mathbf{\omega}_{t+1}^* \in \mathbb{R}^N$ that maximize risk-adjusted returns. We achieve this mapping via an Entropy Pooling framework, which we detail in Section~\ref{sec:integration}.
 We map the qualitative ranking $y_{t+1}$ into continuous portfolio weights $\mathbf{\omega}_{t+1}^* \in \mathbb{R}^N$ through EP (Section~\ref{sec:integration}).

\subsection{Our Two-Stage Training Approach} 

% Our method consists of two-stage training. The first stage is SFT. We fine-tune the model using a high quality CoT reasoning dataset. The second stage is Financial Asset Ranking via Group Relative Policy Optimization (FinRankGRPO), a financial adaptation of GRPO that uses a Spearman-correlation reward for listwise asset ranking. The whole framework is shown in Figure~\ref{fig:main}.

\begin{figure}[htbp] 
\centering 
\includegraphics[width=\linewidth]{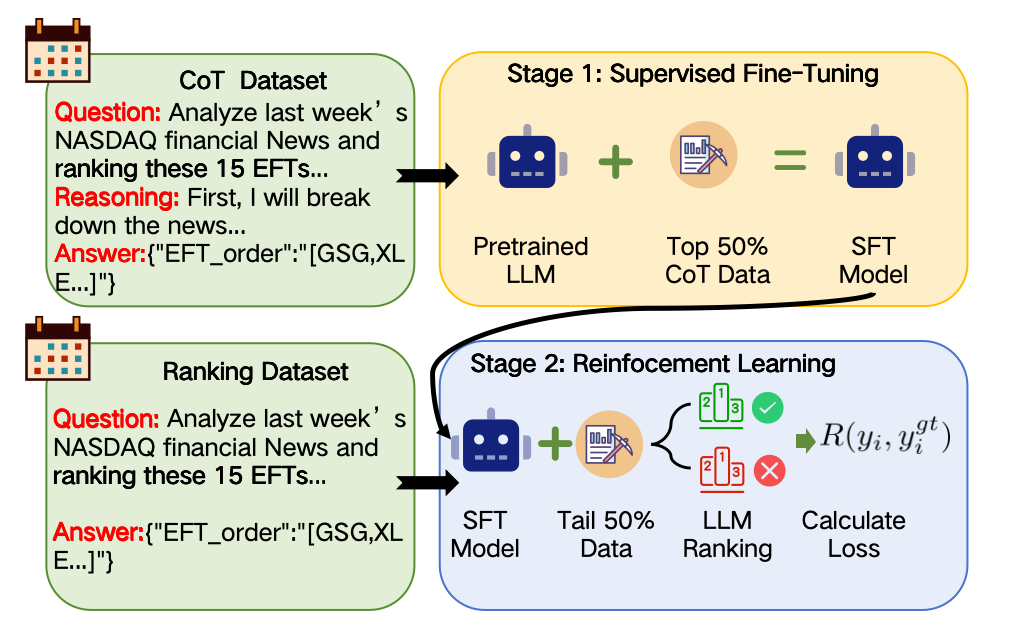} 
\caption{Overview of the two-stage FinRankGRPO pipeline. Stage 1 performs supervised fine-tuning on distilled chain-of-thought ranking data. Stage 2 optimizes the model using a Spearman-based GRPO reward.}

%The two-stage construction framework of our FinRankGRPO, Stage 1 is SFT in high quality CoT datasets, Stage 2 is trained by our FinRankGRPO, which is a Spearman-related ranking reward.}
\label{fig:main}
\end{figure}

Our framework follows a two-stage training procedure. In the first stage, we perform supervised fine-tuning (SFT) on a high-quality chain-of-thought reasoning dataset. In the second stage, we apply Financial Asset Ranking via Group Relative Policy Optimization (FinRankGRPO), our financial adaptation of GRPO that uses a Spearman correlation reward to optimize listwise asset ranking. The overall framework is illustrated in Figure~\ref{fig:main}.

\subsubsection{Stage 1 Supervised Fine-Tuning}

 % The first stage aims to equip the LLM with fundamental reasoning abilities through SFT on high-quality distilled data. We employ commercial models to distill CoT datasets based on financial news summaries. These news summaries are then used as training targets with a standard cross-entropy loss function. The loss for sentence pairs $(x,y)$ is formalized as:

The SFT stage teaches the model to produce structured financial rankings from distilled CoT data generated by commercial models. For each prompt-output pair $(x,y)$, we optimize:

\begin{equation}
\begin{aligned}
\mathcal{L}_{\mathrm{SFT}}(\theta)
&= -\sum_{t=1}^{|y|}
\log \pi_\theta\!\left(y_t \mid x,y_{<t}\right),
\label{sft_equation}
\end{aligned}
\end{equation}
% where $x$ denotes the input prompt and $y$ represents the target reasoning output. During training, a fixed template $\mathcal{I}$ (detailed in Appendix~\ref{appx:prompt}) guides the model to generate structured asset rankings. This stage provides the model with a strong initial capability for producing well-formed rankings.

where $x$ is the input prompt, $y=(y_1,\ldots,y_{|y|})$ is the target response, and $y_{<t}$ denotes its prefix before token $t$. The fixed prompt template is shown in Appendix~\ref{appx:prompt}.

\subsubsection{Stage 2 FinRankGRPO}

% Although the SFT model improves reasoning capabilities, it still struggles to generate high-quality rankings. To address this issue, the second stage introduces FinRankGRPO, which redefines the process of ranking portfolios as a reinforcement learning problem with ranking optimization objectives. 

% The reward is based as the Spearman correlation between the model's predicted ranking and the true asset ranking. We choose  Spearman over other methods for two key reasons: Firstly, unlike the Pearson correlation, which measures linear dependence on raw values and is sensitive to numerical scale, the Spearman correlation evaluates rank order independently of absolute magnitudes. Secondly, whereas metrics such as NDCG place greater importance on the top ranks, Spearman uniformly penalizes ranking errors across the entire list. This makes it better suited to obtaining a globally reliable ordering, which is critical in portfolio construction. 

% As a specialized adaptation of GRPO~\cite{Shao2024DeepSeekMathPT} for listwise financial asset ranking, FinRankGRPO directly optimizes the model via reward policy gradients, effectively aligning its generative outputs with explicit financial asset ranking objectives. For each training instance with input \(x\), FinRankGRPO first samples a group of candidate outputs \(\{o_g\}_{g=1}^{G}\) from the current policy \(\pi_\theta\). A rule-based extraction method \(\mathcal{E}\) is then applied to each \(o_g\) to obtain a structured ranking list \(y_g\). The sampling and extraction procedure is formalized as:

Although the SFT model improves reasoning capabilities, it still struggles to generate high quality rankings. To address this issue, the second stage introduces FinRankGRPO, which redefines the process of ranking portfolios as a reinforcement learning problem with ranking optimization objectives. 

The reward is based on the Spearman correlation between the model's predicted ranking and the true asset ranking. We choose  Spearman over other methods for two key reasons. Firstly, unlike the Pearson correlation, which measures linear dependence on raw values and is sensitive to numerical scale, the Spearman correlation evaluates rank order independently of absolute magnitudes. Secondly, whereas metrics such as NDCG place greater importance on the top ranks, Spearman uniformly penalizes ranking errors across the entire list. This makes it better suited to obtaining a globally reliable ordering, which is critical in portfolio construction.

As a specialized adaptation of GRPO~\cite{Shao2024DeepSeekMathPT} for listwise financial asset ranking, FinRankGRPO directly optimizes the model via reward policy gradients, effectively aligning its generative outputs with explicit financial asset ranking objectives. For each training instance with input \(x\), FinRankGRPO first samples a group of candidate outputs \(\{o_g\}_{g=1}^{G}\) from the current policy \(\pi_\theta\). A rule-based extraction method \(\mathcal{E}\) is then applied to each \(o_g\) to obtain a structured ranking list \(y_g\). The sampling and extraction procedure is formalized as:

\begin{equation}
\begin{aligned}
&\{o_g\}_{g=1}^{G} \sim \pi_\theta(\cdot \mid x), \\
&y_g = \mathcal{E}(o_g), \quad g = 1, \dots, G.
\end{aligned}
\label{eq:sampling_extraction}
\end{equation}

% The quality of each generated ranking \(y_g\) is evaluated by comparing it with the ground-truth ranking \(y^{\text{gt}}\), yielding a ranking-specific reward \(R(y_g, y^{\text{gt}})\). This reward function emphasizes correct ordering and penalizes structural errors:

Each ranking is scored against the ground-truth ordering \(y^{\text{gt}}\). Let \(\rho_g=\rho_s(y_g,y^{\text{gt}})\) denote their Spearman correlation:

\begin{equation}
R(y_g, y^{\text{gt}}) =
\begin{cases}
-1, & \text{if {Error}}; \\[0.5em]
\sigma(\rho_g), & \text{if } \rho_g > 0; \\[0.5em]
\rho_g, & \text{if } \rho_g \leq 0,
\end{cases}
\label{eq:reward}
\end{equation}

% where $\rho_s(y_g, y^{\text{gt}})$ denotes the Spearman rank correlation coefficient between the generated and ground-truth lists, and $\sigma(\cdot)$ is the sigmoid function used to amplify positive rewards. An Error is strictly defined as a structural mismatch: it occurs if $y_g$ contains any asset identifier not present in $y^{\text{gt}}$, or if it omits any required asset. Such errors receive a fixed penalty of $-1$ to strongly discourage hallucinations or omissions.

% The standardized advantage for each response is then computed based on the group’s reward distribution:
% where $\sigma(z)=(1+\exp(-z))^{-1}$ is the standard logistic sigmoid implemented in our training code. Consequently, a valid positive correlation in $(0,1]$ is mapped to $(0.5,0.731]$, whereas a non-positive correlation is left unchanged. This asymmetric mapping creates a positive margin between correctly and incorrectly ordered samples. Structural errors, including empty rankings, duplicated identifiers, missing assets, or hallucinated assets, receive a fixed penalty of $-1$.
where $\sigma(z)=(1+\exp(-z))^{-1}$ is the standard logistic sigmoid implemented. This asymmetric mapping creates a positive margin between correctly and incorrectly ordered samples. Structural errors, including empty rankings, duplicated identifiers, missing assets, or hallucinated assets, receive a fixed penalty of $-1$.

The standardized advantage for each response is then computed based on the group's reward distribution:
\begin{equation}
\hat{A}_g = \frac{R(y_g, y^{\text{gt}}) - \mu_R}
{\sigma_R + \epsilon_A},
\label{eq:advantage}
\end{equation}
where $\mu_R$ and $\sigma_R$ are the mean and standard deviation of the $G$ rewards for the same prompt. The small constant $\epsilon_A>0$ prevents division by zero when all sampled responses receive the same reward. The response-level advantage $\hat{A}_g$ is shared by all generated tokens in response $o_g$.

Finally, the policy is optimized using the following objective:

\begin{equation}
\begin{aligned}
\mathcal{J}_{\mathrm{FR}}(\theta)
&=\mathbb{E}\!\left[\mathcal{L}_{G}(\theta)\right],\\
\mathcal{L}_{G}(\theta)
&=\frac{1}{G}\sum_{g=1}^{G}\frac{1}{|o_g|}
\sum_{t=1}^{|o_g|}\ell_{g,t}(\theta),\\
\ell_{g,t}(\theta)
&=\min\!\left(r_{g,t}(\theta)\hat{A}_g,
\bar r_{g,t}(\theta)\hat{A}_g\right)\\
&\quad-\beta D_{g,t}^{\mathrm{KL}}(\theta).
\end{aligned}
\label{eq:finrankgrpo_objective}
\end{equation}
where the expectation is over training prompts and response groups sampled from $\pi_{\theta_{\mathrm{old}}}$.

Here, the token-level importance ratio and its clipped version are
\begin{equation}
\begin{aligned}
r_{g,t}(\theta)
&=\frac{\pi_\theta(o_{g,t}\mid s_{g,t})}
{\pi_{\theta_{\mathrm{old}}}(o_{g,t}\mid s_{g,t})},\\
\bar r_{g,t}(\theta)
&=\operatorname{clip}\!\left(r_{g,t}(\theta),1-\epsilon,1+\epsilon\right).
\end{aligned}
\label{eq:importance_ratio}
\end{equation}
Here $s_{g,t}=(x,o_{g,<t})$ is the token context. The policy $\pi_{\theta_{\mathrm{old}}}$ is used to sample the response group, and the clipping radius $\epsilon$ limits each update. At token position $t$, $D_{g,t}^{\mathrm{KL}}(\theta)$ denotes $D_{\mathrm{KL}}\!\left(\pi_\theta(\cdot\mid s_{g,t})\,\|\,\pi_{\mathrm{ref}}(\cdot\mid s_{g,t})\right)$, where $\pi_{\mathrm{ref}}$ is the fixed reference policy and $\beta\geq0$ controls the penalty strength. Thus, optimization is performed over the tokens of $o_g$, while the extracted ranking $y_g=\mathcal{E}(o_g)$ is used only to compute the response-level reward. By optimizing a reward derived from rank correlation, FinRankGRPO directly encourages the generated asset ordering to agree with the realized-return ordering.
% which are essential for matching the relative ordering required for portfolio management. Furthermore, by optimizing directly over asset rankings rather than numerical forecasts, we avoid the numerical unreliability inherent in LLM-based point predictions.

\subsection{Integration with Entropy Pooling}
\label{sec:integration}

% To bridge the gap between discrete semantic rankings and continuous portfolio allocation, we employ Entropy Pooling (EP)~\cite{meucci2006beyond, meucci2010fully}. This framework allows us to process the LLM-generated ranking $y_{t+1}$ as a set of relative views, updating the market's prior distribution to a posterior that reflects these semantic insights. The process consists of three steps: view construction, probability update, and portfolio optimization.

To bridge the gap between discrete semantic rankings and continuous portfolio allocation, we use EP~\cite{meucci2006beyond, meucci2010fully} to convert LLM generated rankings into relative views and update the market prior into a posterior distribution used for portfolio construction.

\subsubsection{Constructing View Constraints}
% The ranking list $y_{t+1}$ implies a structured set of relative performance expectations. Specifically, if the LLM ranks asset $i$ higher than asset $j$ (denoted as $i \succ j$), it implies the view that $\mathbb{E}[R_i] \geq \mathbb{E}[R_j]$. We aggregate these pairwise comparisons into a linear inequality system:
If the LLM ranks asset $i$ above asset $j$ (\(i \succ j\)), we impose $\mathbb{E}[R_i] \geq \mathbb{E}[R_j]$. The ranking is encoded as:

\begin{equation}
\mathbf{P} \mathbb{E}_{\mathbf{q}}[\mathbf{R}] \geq \mathbf{v},
\label{eq:constraints}
\end{equation}
% where $\mathbf{R} \in \mathbb{R}^N$ is the vector of asset returns, $\mathbf{P} \in \mathbb{R}^{(N-1) \times N}$ is the ``pick matrix'' identifying the compared assets, and $\mathbf{v} \in \mathbb{R}^{N-1}$ is the view vector (set to $\mathbf{0}$ for qualitative rankings).
where $\mathbf{R}$ is the asset-return vector, $\mathbf{P}$ is the pick matrix, and $\mathbf{v}=\mathbf{0}$ for qualitative rankings. An example is provided in Appendix~\ref{appx:ep_example}.

% \noindent\textbf{Example Construction.} 
% Consider a simplified universe of $N=3$ assets: XLF (Financials), GLD (Gold), and XLE (Energy). If the LLM generates the ranking $\text{XLF} \succ \text{GLD} \succ \text{XLE}$, this implies two views: $\mathbb{E}[R_{XLF}] \geq \mathbb{E}[R_{GLD}]$ and $\mathbb{E}[R_{GLD}] \geq \mathbb{E}[R_{XLE}]$. The corresponding constraint system is defined as:
% \begin{equation}
% \underbrace{\begin{bmatrix} 
% 1 & -1 & 0 \\ 
% 0 & 1 & -1 
% \end{bmatrix}}_{\mathbf{P}} 
% \begin{bmatrix} 
% \mathbb{E}[R_{XLF}] \\ \mathbb{E}[R_{GLD}] \\ \mathbb{E}[R_{XLE}] 
% \end{bmatrix} 
% \geq 
% \underbrace{\begin{bmatrix} 
% 0 \\ 0 
% \end{bmatrix}}_{\mathbf{v}}.
% \end{equation}
% This matrix $\mathbf{P}$ effectively encodes the ordinal information from the LLM into a format ingestible by the optimization engine.

\subsubsection{Minimizing Relative Entropy}
We estimate a posterior scenario distribution $\mathbf{q}^*$ that satisfies Eq.~(\ref{eq:constraints}) while minimizing KL divergence from the empirical prior $\mathbf{p}$:

\begin{equation}
\begin{aligned}
\mathbf{q}^* = \mathop{\arg\min}_{\mathbf{q}} & \sum_{k=1}^{K} q_k \ln \left( \frac{q_k}{p_k} \right) \\
\text{s.t.} \quad & \mathbf{P} \mathbb{E}_{\mathbf{q}}[\mathbf{R}] \geq \mathbf{v}, \\
& \sum_{k=1}^{K} q_k = 1, \quad \mathbf{q} \geq \mathbf{0}.
\end{aligned}
\label{eq:entropy_min}
\end{equation}
% By solving this optimization problem, we obtain a posterior distribution that naturally incorporates the non-linear, semantic insights from the LLM without making parametric assumptions about return distributions.
This posterior incorporates semantic views without parametric assumptions on returns.

\subsubsection{Portfolio Optimization}
% Finally, we utilize the posterior moments derived from $\mathbf{q}^*$, specifically the expected returns $\boldsymbol{\mu}_{\text{EP}} = \mathbb{E}_{\mathbf{q}^*}[\mathbf{R}]$ and covariance $\boldsymbol{\Sigma}_{\text{EP}} = \text{Cov}_{\mathbf{q}^*}[\mathbf{R}]$, to construct the optimal portfolio. The risk-free rate $r$ is set to be 2.5\%. We adopt the Maximum Sharpe Ratio (MSR) objective:
Finally, we use posterior moments $\boldsymbol{\mu}_{\text{EP}}=\mathbb{E}_{\mathbf{q}^*}[\mathbf{R}]$ and $\boldsymbol{\Sigma}_{\text{EP}}=\text{Cov}_{\mathbf{q}^*}[\mathbf{R}]$ to solve a long-only Maximum Sharpe Ratio problem with risk-free rate $r=2.5\%$:

% \begin{equation}
% \mathbf{\omega}^* = \mathop{\arg\max}_{\mathbf{\omega}} \frac{\mathbf{\omega}^\top \boldsymbol{\mu}_{\text{EP}} - r}{\sqrt{\mathbf{\omega}^\top \boldsymbol{\Sigma}_{\text{EP}} \mathbf{\omega}}} \quad \text{s.t.} \quad \mathbf{1}^\top \mathbf{\omega} = 1, \quad \mathbf{\omega} \geq \mathbf{0}.
% \label{eq:msr_final}
% \end{equation}

\begin{equation}
\begin{aligned}
\mathbf{\omega}^*
&= \mathop{\arg\max}_{\mathbf{\omega}}
\frac{\mathbf{\omega}^\top \boldsymbol{\mu}_{\text{EP}} - r}
{\sqrt{\mathbf{\omega}^\top \boldsymbol{\Sigma}_{\text{EP}}\mathbf{\omega}}} \\
\text{s.t.}\quad
&\mathbf{1}^\top \mathbf{\omega} = 1,\qquad
\mathbf{\omega} \geq \mathbf{0}.
\end{aligned}
\label{eq:msr_final}
\end{equation}

The resulting weights $\mathbf{\omega}^*$ dynamically concentrate capital on assets where the LLM predicts high relative outperformance, scaled by the risk estimates refined via EP. Further implementation details of Entropy Pooling and portfolio optimization are provided in Appendix~\ref{appx: details_portfolio_optimization}.

\section{Experiment}

\subsection{Datasets}

We construct a backtesting dataset spanning 2010--2025 over 15 highly liquid ETFs, including size-based indices, GICS sectors, and alternative assets. Market prices are obtained from Bloomberg\footnote{\url{https://www.bloomberg.com/}} under our institution's academic license. Financial news headlines are collected from FNSPID~\cite{Dong2024FNSPIDAC}, released under the CC BY-NC 4.0 license, for 2010--2023 and from The Wall Street Journal archive for 2024--2025\footnote{\url{https://www.wsj.com/news/archive/years}}.

We use a chronological split to avoid leakage from task-specific training into evaluation: data from 2010--2019 are used for training, and data from 2020--2025 are reserved for out-of-sample testing. The training set contains 2,516 instances constructed using rolling one-week windows, whereas the test set contains 314 non-overlapping calendar-week instances. This distinction provides sufficient training examples while ensuring that the reported evaluation is conducted on non-overlapping weekly periods.

For each training instance, we collect candidate ranking responses from GPT-4o-mini~\cite{openai2024gpt4technicalreport}, Claude-3.5-Sonnet~\cite{anthropic2024claude3}, and DeepSeek-Chat~\cite{deepseekai2025deepseekv32}, and retain the response with the highest Spearman correlation. Further details on teacher-response selection and dataset partitioning are provided in Appendix~\ref{appx:dataset}.

\subsection{Baselines}

We evaluate our proposed method using two distinct categories of metrics, including rank-based metrics and financial performance metrics.

Traditional Finance Methods: Equal Weights (EW) is a non-parametric heuristic that allocates capital uniformly across all $N$ assets ($w_i = 1/N$).
{Mean-Variance Optimization (MVO-SR)} is the classical Markowitz framework. We adopt the Maximum Sharpe Ratio (SR) variant.
{Short-Term Momentum (Momentum)} is a trend following strategy that ranks assets based on their realized returns over the preceding two weeks. 

LLM-Based Methods: For commercial models, we evaluate via their APIs, including GPT-4o-mini~\cite{openai2024gpt4technicalreport}, Claude-3.5-Sonnet (20241022)~\cite{anthropic2024claude3}, and DeepSeek-Chat (V3.2)~\cite{deepseekai2025deepseekv32}.
Appendix~\ref{appx:other_methods} additionally evaluates DeepSeek-V3.2 with one-shot list generation and point-wise return prediction.
% together with FinBERT--RankNet and LambdaMART non-LLM rankers, under the same ETF universe and backtest protocol.
% For the model-scale analysis, we use Qwen2.5 checkpoints at 0.5B, 1.5B, and 3B parameters, including Base and Instruct variants. The main experiments use Qwen3-4B and Qwen3-4B-Instruct.
For open source models, we use Qwen3-4B-Base\footnote{\url{https://huggingface.co/Qwen/Qwen3-4B-Base}}and Qwen3-4B-Instruct\footnote{\url{https://huggingface.co/Qwen/Qwen3-4B-Instruct-2507}} as the backbone models.
% These models serve as the foundation for our SFT and subsequent FinRankGRPO stages.

Additional comparisons with FinBERT, RankNet, and LambdaMART are provided in Appendix~\ref{appx:other_methods}.

\subsection{Evaluation}

We evaluate our method using ranking metrics and financial performance metrics. Details of the backtest protocol are provided in the Appendix~\ref{appx:detailed_statistics}.

\textbf{Ranking metrics.}
Ranking metrics assess the agreement between the generated asset ordering and the ground-truth ranking induced by subsequent realized returns. We report Spearman's rank correlation (\(\rho\)), Kendall's tau (\(\tau\)), and normalized discounted cumulative gain at \(k\) (NDCG@\(k\)).

\textbf{Financial performance metrics.}
Financial performance metrics assess the downstream utility of the resulting portfolios. We report annualized return (Ann. Ret.), Sharpe ratio (Sharpe), maximum drawdown (Max DD), and Calmar ratio (Calmar).

% \subsection{Main Results}
% \begin{table*}[htbp]
% \centering
% \small
% \setlength{\tabcolsep}{2pt} % 减少列间距
% % \begin{tabular}{@{}l|cccc|cccc@{}}
% \begin{tabularx}{\linewidth}{@{}l|*{4}{X}|*{4}{X}@{}}
% \toprule
% \multirow{2}{*}{\textbf{Model}} & \multicolumn{4}{c|}{\textbf{Ranking Metrics}} & \multicolumn{4}{c}{\textbf{Portfolio Performance}} \\
% \cmidrule(lr){2-5} \cmidrule(lr){6-9}
% % & \textbf{Spearman} & \textbf{Kendall's Tau} & \textbf{NDCG@3} & \textbf{NDCG@5} &  \textbf{Annualized Return}& \textbf{Sharpe} & \textbf{Maximum Drawdown} & \textbf{Calmar Ratio} \\

% & \textbf{Spear.} & \textbf{Kend.} & \textbf{NDCG@3} & \textbf{NDCG@5} & \textbf{Ann. Ret.} & \textbf{Sharpe} & \textbf{Max DD} & \textbf{Calmar} \\

% \midrule
% % \multicolumn{9}{l}{\textit{Ground Truth }} \\
% \multicolumn{9}{p{\linewidth}}{\textit{Ground Truth}} \\   % 改这里

% \midrule
% Ground Truth & 1.000 & 1.000 & 1.000  & 1.000   &1.239 & 6.670 & 0.175 & 7.088 \\
% \midrule
% % \multicolumn{9}{l}{\textit{Traditional Finance Baselines}} \\
% \multicolumn{9}{p{\linewidth}}{\textit{Traditional Finance Baselines}} \\   % 改这里

\subsection{Main Results}
\begin{table*}[htbp]
\centering
\small
\setlength{\tabcolsep}{4pt}
\renewcommand{\arraystretch}{1.08}
\begin{tabularx}{\linewidth}{@{}l*{8}{>{\centering\arraybackslash}X}@{}}
\toprule
\multirow{2}{*}{\textbf{Model}} & \multicolumn{4}{c}{\textbf{Ranking Metrics}} & \multicolumn{4}{c}{\textbf{Portfolio Performance}} \\
\cmidrule(lr){2-5} \cmidrule(lr){6-9}
& \textbf{Spear.} & \textbf{Kend.} & \textbf{N@3} & \textbf{N@5} & \textbf{Ann. Ret.} & \textbf{Sharpe} & \textbf{Max DD} & \textbf{Calmar} \\
\midrule
% \multicolumn{9}{@{}l}{\textit{Ground Truth}} \\
Ground-truth ranking + EP & 1.000 & 1.000 & 1.000 & 1.000 & 1.049 & 7.653 & 0.138 & 7.619 \\
\midrule
\multicolumn{9}{@{}l}{\textit{Traditional Finance Baselines}} \\
EW & -- & -- & -- & -- & \underline{0.119} & 0.502 & 0.356 & 0.334 \\
MVO-SR & -0.011 & -0.009 & 0.567 & 0.595 & 0.094 & 0.428 & 0.268 & 0.349 \\
Momentum & -0.027 & -0.019 & 0.544 & 0.581 & 0.058 & 0.238 & \underline{0.209} & 0.280 \\
\midrule
\multicolumn{9}{@{}l}{\textit{LLM Prompting-Based Methods}} \\
DeepSeek-Chat & 0.015 & \textbf{0.016} & 0.574 & \textbf{0.605} & 0.108 & 0.560 & 0.227 & 0.478 \\
GPT-4o-mini & 0.014 & 0.013 & 0.570 & 0.604 & 0.081 & 0.363 & 0.240 & 0.335 \\
Claude & 0.001 & 0.004 & 0.570 & 0.597 & 0.101 & 0.396 & 0.254 & 0.398 \\

\midrule
\multicolumn{9}{@{}l}{\textit{LLM Post-Training-Based Methods}} \\
Qwen3-4B & -0.015 & -0.010 & \textbf{0.601} & 0.582 & 0.070 & 0.286 & \textbf{0.161} & 0.389 \\
+SFT & 0.012 & 0.010 & 0.564 & 0.596 & 0.109 & 0.558 & 0.236 & 0.467 \\
+FinRankGRPO & \underline{0.015} & 0.013 & \underline{0.575} & 0.602 & 0.111 & \underline{0.574} & 0.220 & \underline{0.512} \\

\midrule
Qwen3-4B-Instruct & 0.009 & 0.007 & 0.575 & \underline{0.604} & 0.094 & 0.431 & 0.260 & 0.362 \\
+SFT & 0.014 & 0.012 & 0.572 & 0.601 & 0.107 & 0.532 & 0.222 & 0.484 \\
+FinRankGRPO & \textbf{0.018} & \underline{0.015} & 0.571 & 0.602 & \textbf{0.131} & \textbf{0.689} & 0.210 & \textbf{0.625} \\

\bottomrule
\end{tabularx}

\caption{Main experimental results comparing traditional finance baselines, prompting-based commercial LLMs, and post-trained open-weight models. Ground-truth ranking + EP is a non-deployable oracle that supplies the realized return order to the same Entropy Pooling (EP) and portfolio pipeline. Bold indicates the best deployable performance, and underline indicates the second-best deployable performance.  All results represent the mean performance averaged over 5 independent trials.}

\label{tab:main_results}
\end{table*}

Table~\ref{tab:main_results} demonstrates that our proposed FinRankGRPO method, which is finetuned from the Qwen3-4B-Instruct model, achieves the highest portfolio performance. Appendix Tables~\ref{tab:main_stats_ranking} and~\ref{tab:main_stats_portfolio} further report dispersion and 95\% confidence intervals.

\textbf{Ranking performance}. Qwen3-4B-Instruct with FinRankGRPO achieves the highest mean Spearman correlation of $0.018$ among the evaluated methods. Its Kendall's tau of $0.015$ is second only to DeepSeek-Chat with $0.016$, indicating that the proposed training objective obtains the highest mean Spearman correlation in the main comparison while remaining competitive on pairwise ordering.

\textbf{Portfolio outcomes}. FinRankGRPO achieves the highest mean annualized return of $0.131$ and Sharpe ratio of $0.689$, outperforming both the equal-weight benchmark, with an annualized return of $0.119$ and a Sharpe ratio of $0.502$, and the strongest prompting-based commercial model, DeepSeek-Chat, with an annualized return of $0.108$ and a Sharpe ratio of $0.560$. Two-sided paired tests show significant improvements in annualized return and Sharpe ratio relative to standard GRPO, with respective $p$-values of $0.004$ and $0.005$, and relative to DeepSeek-Chat, with respective $p$-values of $0.013$ and $0.026$. Appendix Table~\ref{tab:significance_tests} reports the complete comparisons.

Although Spearman correlation and the Sharpe ratio are not simply linearly related, they tend to move in the same direction. Spearman correlation captures average ordinal agreement, whereas the Sharpe ratio depends on how rank changes are translated into portfolio weights and how those weights interact with realized returns and volatility. Overall, the joint gains in ranking and portfolio performance indicate that ranking-based rewards improve downstream portfolio management performance.

\section{Analysis and Interpretability}

\subsection{The impact of Ranking Metrics}
\begin{table}[htbp]
\centering
\small
\setlength{\linewidth}{2pt}
\renewcommand{\arraystretch}{1.08}
\begin{tabular}{@{}lrrrr@{}}
\toprule
\textbf{Reward Metric} & \textbf{Spear.} & \textbf{Kendall} & \textbf{Sharpe} & \textbf{Max DD} \\
\midrule
NDCG@3 & 0.014 & 0.011 & 0.344 & 0.234 \\
Kendall's Tau & 0.012 & 0.009 & 0.522 & 0.237 \\
GRPO & 0.013 & 0.012 & 0.436 & 0.249 \\
FinRankGRPO & \textbf{0.018} & \textbf{0.015} & \textbf{0.689} & \textbf{0.210} \\
\bottomrule
\end{tabular}
\caption{Comparison of reward metrics using the same SFT-initialized Qwen3-4B-Instruct model.}
\label{tab:ranking_metric_reward}
\end{table}

To examine the effect of reward design, we compare the Spearman-based FinRankGRPO objective with three alternatives: NDCG@3, Kendall's tau, and the exact-match reward used in our standard-GRPO variant. All variants are initialized from the same SFT checkpoint. The comparison therefore evaluates not only the final ranking performance associated with each reward, but also whether the reward provides an informative learning signal during policy optimization.

As shown in Table~\ref{tab:ranking_metric_reward}, FinRankGRPO achieves the highest Spearman correlation and Sharpe ratio among the evaluated variants. Compared with standard GRPO, it increases the Spearman correlation from $0.013$ to $0.018$ and the Sharpe ratio from $0.436$ to $0.689$. This improvement is accompanied by a substantially denser optimization signal. In the examined training trace, the exact-match reward assigns zero reward to every sampled completion, resulting in zero advantage and, consequently, no effective policy update. By contrast, approximately $99.5\%$ of the completions sampled by FinRankGRPO receive non-zero rewards, and $99.5\%$ of its optimization steps produce a non-zero training loss. These observations indicate that a graded ranking reward can distinguish partially correct orderings even when the generated ranking does not exactly match the target.

The three ranking-based rewards differ in the aspects of the generated ordering that they emphasize. NDCG@3 concentrates primarily on the highest-ranked assets and gives comparatively limited consideration to errors lower in the list. Kendall's tau evaluates pairwise ordering consistency, treating each correctly or incorrectly ordered pair as the basic unit of comparison. Spearman correlation instead measures agreement between the predicted and target ranks across the complete list, providing graded feedback according to the magnitude of rank deviations. This property is particularly suitable for our setting because assets throughout the ranking contribute to the relative-view constraints used by Entropy Pooling. Consequently, the Spearman-based reward aligns the optimization signal more closely with the downstream use of the generated rankings.

% \input{table/2.ranking_metrics}

% This is because portfolio construction requires a globally coherent ordering over the full asset universe, rather than only accurate top-ranked assets. NDCG@3 mainly emphasizes the first few positions and provides weak supervision for the middle and lower parts of the list, even though these assets still affect the relative-view constraints used by Entropy Pooling. Kendall's tau measures pairwise order consistency, but its pairwise counting objective can be less informative as a dense training signal because it treats local inversions uniformly and does not directly optimize smooth global rank correlation. In contrast, Spearman correlation evaluates the full list, provides a continuous ordinal reward, and better aligns the model with stable relative asset preferences under noisy financial news.

\subsection{The Impact of Two-Stage Training}

\begin{figure}[htbp]
    \centering
    \includegraphics[width=\linewidth]{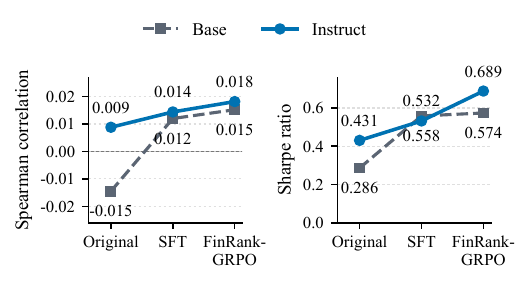}
    \caption{Post-training improves both Qwen3-4B variants, and the Instruct FinRankGRPO checkpoint performs best. Left: Spearman correlation. Right: Sharpe ratio.}
    \label{fig:two_stage_training_impact_3B}
\end{figure}

Figure~\ref{fig:two_stage_training_impact_3B} shows that SFT improves both ranking and portfolio performance for the Qwen3-4B Base and Instruct models, while FinRankGRPO provides further gains over the corresponding SFT checkpoints. The complete pipeline achieves its best performance with the Instruct model, reaching a Spearman correlation of $0.018$ and a Sharpe ratio of $0.689$. Overall, SFT improves output validity and mean ranking metrics, while FinRankGRPO further improves the reported means under the evaluated protocol.

\subsection{The Ablation Study of Input Information}

% To validate that FinRankGRPO captures genuine semantic signals from financial news rather than relying on superficial pattern matching, memorized historical returns, or simple trend following, we conduct an input ablation study, where news or trading date information are removed to quantify their marginal contribution, shown as Figure~\ref{fig:ablationinput}.

To validate that Qwen3-4B-Instruct with FinRankGRPO captures semantic signals from financial news rather than relying on superficial temporal patterns, we remove the trading date, the news content, or both inputs.

\begin{figure}[htbp]
    \centering
    \includegraphics[width=\linewidth]{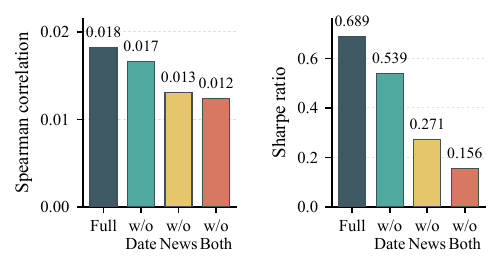}
    \caption{Removing news produces the largest input-ablation loss, showing that the model uses semantic information beyond the trading date. Left: Spearman correlation. Right: Sharpe ratio.}
    \label{fig:ablationinput}
\end{figure}

Figure~\ref{fig:ablationinput} shows that both inputs contribute to performance. Removing the trading date reduces the Sharpe ratio from $0.689$ to $0.539$, while removing the news reduces it further to $0.271$. Removing both inputs yields the weakest result, with a Sharpe ratio of $0.156$ and a Spearman correlation of $0.012$. The larger degradation from removing news confirms that the model relies on semantic market information rather than only temporal cues.

\subsection{Sensitivity to Sentiment Inversion}

To evaluate contextual faithfulness, we conduct a sentiment inversion experiment. We reverse the sentiment polarity of news summaries while preserving entities and event structures, using DeepSeek-Chat as the rewriting model. An example is shown in Appendix~\ref{appx:sentiment}. A faithful model should change its rankings under inverted news, whereas a memorization driven model should remain insensitive to the altered context.

\begin{table}[htbp]
\small
\renewcommand{\arraystretch}{1.08}
\begin{tabularx}{\linewidth}{@{}lXXX@{}}
\toprule
\textbf{Model} &
\multicolumn{1}{c}{\textbf{Original}} &
\multicolumn{1}{c}{\textbf{Inverted}} &
\multicolumn{1}{c}{\textbf{Diff}} \\
\cmidrule(lr){2-2} \cmidrule(lr){3-3} \cmidrule(lr){4-4}
& \textbf{Spear.} & \textbf{Spear.} & \textbf{Spear.} \\
\midrule
Qwen3-4B-Instruct & 0.009 & -0.021 & {0.029} \\
+SFT              & 0.014 & -0.012 & 0.026 \\
+FinRankGRPO      & 0.018 & -0.003 & 0.021 \\
\bottomrule
\end{tabularx}
\caption{Rank correlation under original and sentiment-inverted news prompts (Spearman only). Larger differences indicate greater sensitivity to semantic direction.}
\label{tab:faithfulness_check}
\end{table}

% \begin{table*}[htbp]
% % \centering
% \small
% \setlength{\linewidth}{1pt}
% \renewcommand{\arraystretch}{1.08}
% \begin{tabular}{@{}lcccccc@{}}
% \toprule
% \textbf{Model} &
% \multicolumn{2}{c}{\textbf{Original}} &
% \multicolumn{2}{c}{\textbf{Inverted}} &
% \multicolumn{2}{c}{\textbf{Diff}} \\
% \cmidrule(lr){2-3} \cmidrule(lr){4-5} \cmidrule(lr){6-7}
% & \textbf{Spear.} & \textbf{Kend.} & \textbf{Spear.} & \textbf{Kend.} & \textbf{Spear.} & \textbf{Kend.} \\
% \midrule
% DeepSeek-Chat     & {0.025} & {0.015} & -0.002 & -0.003 & \textbf{0.027} & \textbf{0.018} \\
% GPT-4o-mini  & 0.007 & 0.006 & 0.002 & 0.003 & 0.005 & 0.003 \\
% Claude   & 0.006 & 0.003 & -0.001 & -0.001 & 0.007 & 0.004 \\
% Qwen2.5-3B-Instruct & 0.002 & -0.002 & 0.020 & 0.021 & -0.018 & -0.023 \\
% +SFT   & 0.020 & 0.014 & 0.002 & 0.000 & 0.018 & \underline{0.014} \\
% +FinRankGRPO & {0.026} & {0.018} & 0.006 & 0.005 & \underline{0.020} & 0.013 \\
% \bottomrule
% \end{tabular}
% \caption{Rank correlation under original and sentiment-inverted news prompts. Larger differences indicate greater sensitivity to semantic direction.}
% \label{tab:faithfulness_check}
% \end{table*}

Table~\ref{tab:faithfulness_check} quantifies each model's dependence on input semantics. Qwen3-4B-Instruct with FinRankGRPO changes from $0.018$ Spearman on the original news to $-0.003$ after sentiment inversion, a drop of $0.021$. The corresponding sign reversal indicates that its ranking decisions respond to the semantic polarity of the news.

\subsection{Stress Test: the 2020 Recession Period}

% To evaluate whether our models genuinely understand market risk semantics rather than simply overfitting to historical patterns, we conduct a targeted analysis of the February-April 2020 recession period designated by the National Bureau of Economic Research. This period is characterized by extreme market recession, making it an ideal stress test for evaluating adaptive portfolio construction. We analyze model behavior across ranking metrics, scores, and backtest performance during the recession period.

% Table~\ref{tab:recession_metrics} shows that FinRankGRPO attains the highest Spearman correlation and a Sharpe ratio of $0.486$ during the 2020 recession window. In the 2020-03-09 case study, it ranks Gold (GLD) and Consumer Staples (XLP) near the top, consistent with the defensive signals in the supplied news. This case illustrates model behavior rather than establishing a causal interpretation of the portfolio outcome.
To examine model behavior under an extreme market regime, rather than simply overfitting to historical patterns, we conduct a targeted analysis of the February-April 2020 recession period designated by the National Bureau of Economic Research. This period is characterized by extreme market recession, making it an ideal stress test for evaluating adaptive portfolio construction. We analyze model behavior across ranking metrics, scores, and backtest performance during the recession period.

Table~\ref{tab:recession_metrics} shows that FinRankGRPO attains the highest Sharpe ratio ($0.7954$) and the lowest maximum drawdown ($0.2014$) during the 2020 recession window, while DeepSeek-Chat has the highest Spearman ($0.1195$) and Kendall ($0.0983$) correlations. In the 2020-03-09 case study, FinRankGRPO and SFT both rank Gold (GLD) second, ahead of the base model's third place. 
% This is consistent with a conventional defensive interpretation of gold amid market stress. 
% This case illustrates ranking behavior rather than establishing a causal interpretation of the portfolio outcome.

\begin{table}[htbp]
\centering
\small
\setlength{\tabcolsep}{3pt}
\begin{tabular}{@{}lrrrr@{}}
\toprule
\textbf{Model} &
\textbf{Spear.} & \textbf{Kendall} &
\textbf{Sharpe} & \textbf{Max DD} \\
\midrule
DeepSeek-Chat       & \textbf{0.1195} & \textbf{0.0983} & \underline{0.7457} & 0.2074 \\
% GPT-4o-mini       & 0.1237 & 0.0964 & -0.1128 & 0.2348 \\
% Claude            & 0.0915 & 0.0699 & 0.1455 & 0.2355 \\
Qwen3-4B-Instruct   & 0.0271 & 0.0154 & -0.3592 & 0.2463 \\
+SFT                & 0.0486 & 0.0386 & 0.1192 & \underline{0.2067} \\
+FinRankGRPO        & \underline{0.0893} & \underline{0.0650} & \textbf{0.7954} & \textbf{0.2014} \\
\bottomrule
\end{tabular}
\caption{
Ranking and performance metrics during the February--April 2020 recession period, averaged across five trials per method. 
% Higher values are better for Spearman, Kendall, and Sharpe ratio;
% lower values are better for Max Drawdown, reported as a positive magnitude.
% Bold indicates the best performance among the displayed methods, and underline indicates the second-best performance.
}
\label{tab:recession_metrics}
\end{table}

\section{Conclusion}

% FinRankGRPO reformulates LLM-assisted asset allocation as listwise ranking rather than direct numerical estimation. The generated rankings are integrated with Entropy Pooling, while supervised finetuning and a Spearman-based GRPO stage teach structured output and ordinal alignment. On the ETF benchmark, the best configuration attains the highest mean annualized return and Sharpe ratio among the evaluated methods; the stock transfer study further shows that post-training substantially reduces invalid outputs, although its negative rank correlations delimit the current evidence for cross-asset generalization. Overall, the results support using LLMs as ranking components that complement established quantitative decision mechanisms.
% Future research will explore the extension of this framework to larger asset universes.
We presented FinRankGRPO, a ranking-based framework for LLM-assisted asset allocation. Rather than directly using LLMs to generate numerical estimates, we reformulate their role as extracting relative asset prospects from market news and producing listwise rankings. This design better matches the comparative reasoning strengths of LLMs while avoiding the numerical instability. We further connect the LLM output with Entropy Pooling. FinRankGRPO employs a two-stage training procedure. The first stage uses supervised finetuning on distilled CoT reasoning data to teach structured financial ranking behavior,
while the second stage applies FinRankGRPO with a Spearman-based reward to align generated rankings with realized market orderings. Across the evaluated 15-ETF backtest, FinRankGRPO achieves the highest mean annualized return and Sharpe ratio among the compared methods. Our work contributes a general perspective on adapting LLMs to classical financial decisions, which demonstrates that LLMs can complement established quantitative decision mechanisms.

\section*{Limitations}

This work has two main limitations. First, our experiments are conducted only in English. Although English financial news covers a large portion of market information, it may not fully capture market signals expressed in other languages, especially for region specific assets and local financial events. Extending FinRankGRPO to multilingual financial corpora is an important direction for future work.

Second, computational resource constraints limited our experiments to models with at most 4B parameters. Although the model-scale analysis covers several checkpoints and provides initial evidence about the effect of model capacity, these results may not generalize to substantially larger LLMs. Larger models may exhibit different ranking, reasoning, and optimization behavior under the FinRankGRPO framework. Future work should evaluate FinRankGRPO on larger open-weight models and investigate whether increased model capacity leads to more consistent improvements in ranking quality and portfolio performance.

\section*{Ethical Considerations}

This work uses publicly available financial news and market data for research purposes. We do not use private personal information. The assets and market signals analyzed in this paper are based on open-source financial corpora and publicly observable market outcomes.

We used AI tools to assist with language polishing. All technical ideas, experimental design, results, and conclusions were completed and finalized by the authors.

Our study also constructs synthetic news with opposite market sentiment for analysis. These synthetic examples are used only as controlled research inputs to evaluate model behavior under sentiment perturbations. They are not intended to represent real market events, and should not be used for investment advice, public dissemination, or market manipulation.

\bibliography{custom}

\appendix

\section{Prompt}

\label{appx:prompt}
\begin{figure}[ht]
    % \centering
    \includegraphics[width=\linewidth]{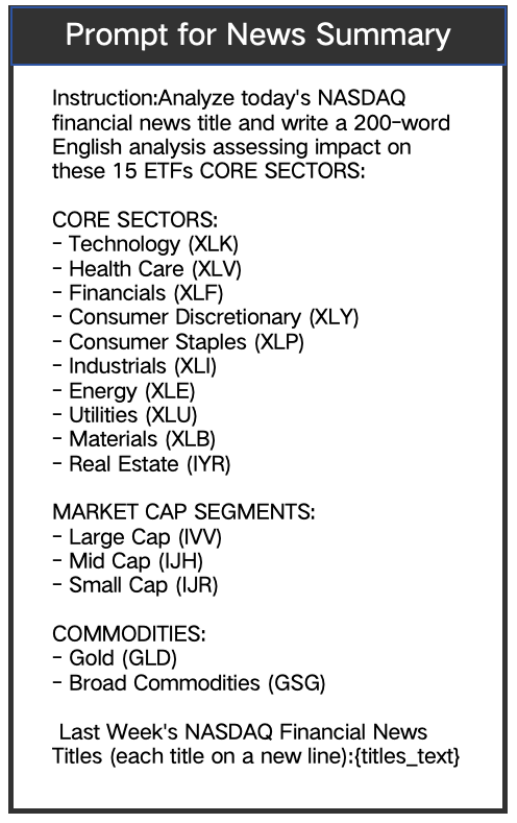}
    \caption{Prompt for Summary News Title.}
    \label{fig:promptsummary}
\end{figure}

As shown in Figure~\ref{fig:promptsummary}, we utilize a summarization prompt that instructs the model to distill a week's worth of raw Nasdaq headlines into a concise set of thematic market signals.

% We design summary follow~\cite{Lee2024FinanceWA}

\begin{figure}
    % \centering
    \includegraphics[width=\linewidth]{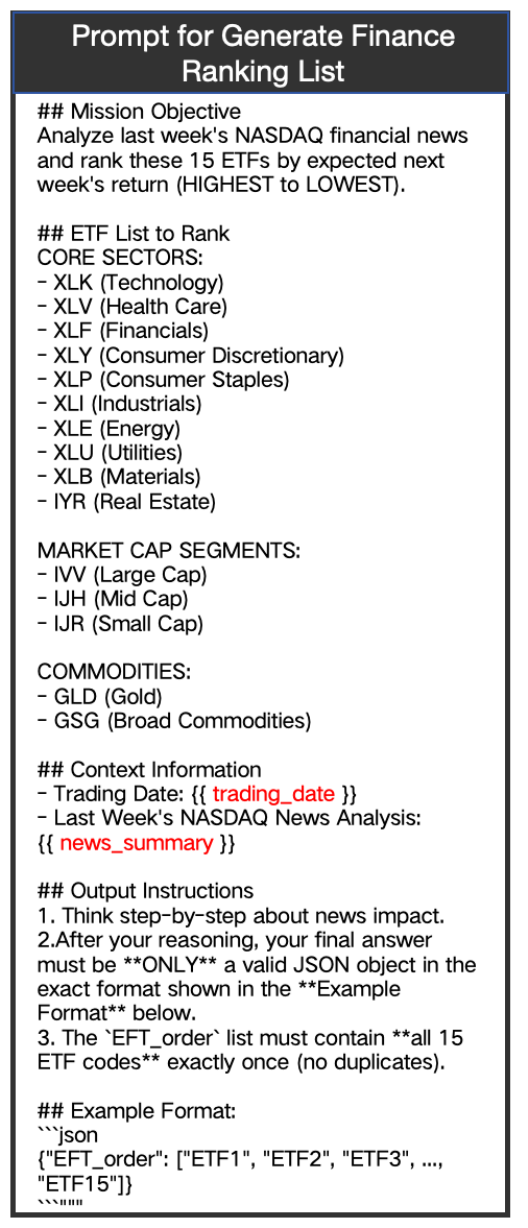}
    \caption{Prompt for Generate Finance Ranking List}
    \label{fig:pormptgenerateranking}
\end{figure}

To generate the training data for the SFT and reinforcement learning stages, we design a prompt that requires the models to return rankings in JSON format, as shown in Figure~\ref{fig:pormptgenerateranking} 

To generate the reverse sentiment news, we design a prompt to invert the sentiment of the original news

\begin{figure}
    % \centering
    \includegraphics[width=\linewidth]{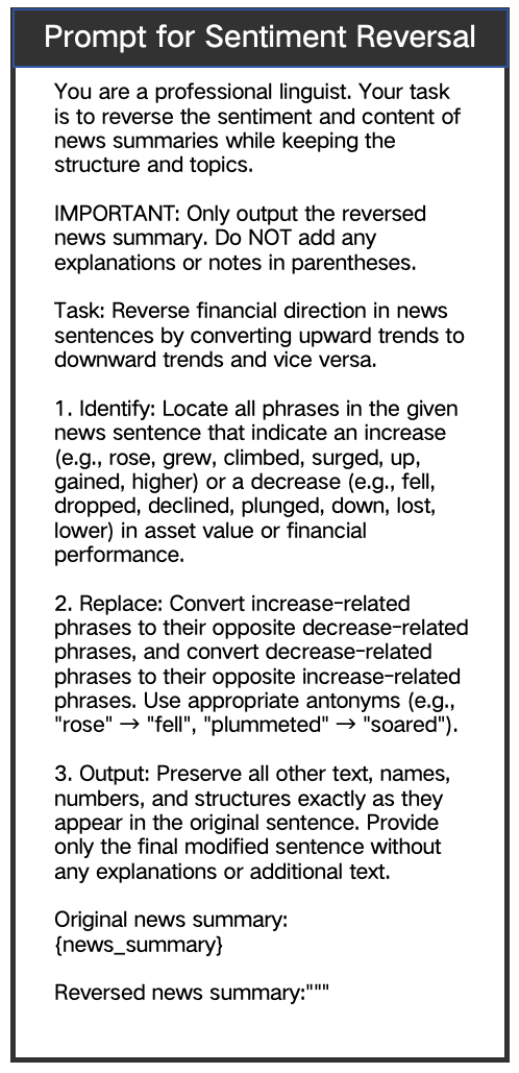}
    \caption{Prompt for Sentiment Inversion Protocol}
    \label{fig:promptforrevers}
\end{figure}

\section{Dataset Details}
\label{appx:dataset}

Our investment universe contains 15 liquid ETFs that represent broad asset classes rather than individual stocks. This design makes the task closer to institutional asset allocation and requires the model to connect macroeconomic narratives with sector and asset class level implications. The full universe is shown in Table~\ref{tab:asset_pool}.

\begin{table}[h]
\centering
\small
\begin{tabularx}{\linewidth}{@{}lX@{}}
\toprule
\textbf{Category} & \textbf{Tickers} \\
\midrule
Indices (Size) & IVV (S\&P 500), IJH (Mid-Cap), IJR (Small-Cap) \\
Sectors & XLB (Materials), XLE (Energy), XLF (Financials), XLI (Industrials), XLK (Information Technology), XLP (Consumer Staples), XLU (Utilities), XLV (Health Care), XLY (Consumer Discretionary) \\
Alternatives & GLD (Gold), GSG (Commodities), IYR (Real Estate) \\
\bottomrule
\end{tabularx}
\caption{Investment universe of 15 ETFs.}
\label{tab:asset_pool}
\end{table}

% Dataset construction consists of four steps. First, we obtain ETF market prices from Bloomberg and collect financial news headlines from two public sources: FNSPID~\cite{Dong2024FNSPIDAC} for 2010--2023 and The Wall Street Journal archive for 2024--2025. For each trading day, headlines from the preceding week are aggregated, then duplicated with SimHash using a Hamming distance threshold of 25, final summarized into asset focused market digests using DeepSeek-Chat. The summarization prompt is shown in Appendix~\ref{appx:prompt}.

Dataset construction consists of four steps. First, we obtain ETF market prices from Bloomberg and collect financial news headlines from two public sources: FNSPID~\cite{Dong2024FNSPIDAC} for 2010--2023 and \textit{The Wall Street Journal} archive for 2024--2025. For each trading week, we aggregate headlines from the preceding week, deduplicate them using SimHash with a Hamming distance threshold of 25.

Second, we use a chronological split to reduce look-ahead bias: 2010--2019 for training and 2020--2025 for out-of-sample testing. The training set contains 2,516 weekly samples, and the test set contains 314 weekly samples. Training samples are produced with a rolling weekly window, while test samples are non-overlapping calendar weeks to simulate realistic out of sample test.

Third, we generate CoT ranking data with GPT-4o-mini~\cite{openai2024gpt4technicalreport}, Claude-3.5-Sonnet~\cite{anthropic2024claude3}, and DeepSeek-Chat~\cite{deepseekai2025deepseekv32}. Each output must follow the specified JSON format. Invalid outputs are discarded and replaced with an empty list. For each weekly sample, we retain the response with the highest Spearman correlation against the ground-truth return ranking. Finally, to support the two-stage pipeline, samples within each calendar year are sorted by Spearman correlation. The top 50\% form the SFT subset, and the bottom 50\% form the more challenging FinRankGRPO subset. This yearly partition preserves exposure to different market regimes in both stages.

\section{Stock Universe Asset}
\label{appx:stock_transfer}

We further evaluate whether the ranking-and-EP pipeline transfers from ETFs to individual stocks. The stock universe comprises 24 U.S. equities spanning all 12 Fama--French industry groups, with two large-cap firms with long price histories selected from each group. The resulting universe includes KO, PM, PCAR, F, GE, CAT, XOM, CVX, PG, SHW, NVDA, AAPL, NFLX, TMUS, NEE, SO, AMZN, WMT, LLY, JNJ, JPM, BRK, V, and MA. Candidate stocks are ranked by market capitalization within each industry, and the final universe is restricted to firms with sufficient historical coverage for the experiment.

The chronological protocol follows the main study. The training pool contains 2,516 weekly observations from 2010--2019, while the out-of-sample test set contains 314 weekly observations from 2020--2025. News is matched only to the corresponding trading date. Within the training period, teacher-generated rankings are evaluated against realized-return orderings. The upper half of the samples is used for SFT, and the lower half is used for FinRankGRPO. 

The stock prompt provides the trading date, a news summary, and the complete list of candidate stocks. The model is instructed to rank all 24 stocks by expected next-period return, explain the relevant company-specific, peer, industry, and macroeconomic effects, and conclude with exactly one JSON object whose \texttt{ticker\_order} field contains every supplied ticker exactly once. This setting changes the asset universe while preserving the listwise prediction format used in the main experiments.

\begin{table}[t]
\centering
\scriptsize
\setlength{\tabcolsep}{3.0pt}
\renewcommand{\arraystretch}{1.08}
\begin{tabular}{@{}lrrrrr@{}}
\toprule
\multicolumn{6}{@{}l}{\textit{Ranking metrics}} \\
\midrule
\textbf{Model} & \textbf{Spear.} & \textbf{Kend.} & \textbf{N@3} & \textbf{N@5} & \textbf{Null} \\
\midrule
Ground Truth & 1.000 & 1.000 & 1.000 & 1.000 & 0.000 \\

% EW & -- & -- & -- & -- & -- \\
% MVO-SR & 0.012 & 0.009 & 0.567 & 0.577 & 0.000 \\
% Momentum & 0.006 & 0.005 & 0.552 & 0.564 & 0.000 \\
DeepSeek-Chat & -0.024 & -0.016 & 0.524 & 0.545 & 0.115 \\
\midrule
% \midrule
Qwen3-4B-Instruct & -0.145 & -0.105 & 0.457 & 0.520 & 0.994 \\
+SFT & -0.024 & -0.016 & 0.534 & 0.547 & \textbf{0.052} \\
+FinRankGRPO & \textbf{-0.017} & -0.012 & \textbf{0.537} & \textbf{0.551} & 0.106 \\
\bottomrule
\end{tabular}

\vspace{3pt}
\begin{tabular}{@{}lrrrr@{}}
\toprule
\multicolumn{5}{@{}l}{\textit{Portfolio metrics}} \\
\midrule
\textbf{Model} & \textbf{Ann. Ret.} & \textbf{Sharpe} & \textbf{Max DD} & \textbf{Calmar} \\
\midrule
Ground Truth & 3.790 & 16.026 & 0.124 & 30.604 \\
% \midrule
% EW & 0.218 & 1.007 & 0.319 & 0.684 \\
% MVO-SR & 0.201 & 0.731 & 0.320 & 0.627 \\
% Momentum & 0.123 & 0.494 & 0.300 & 0.412 \\
DeepSeek-Chat & 0.139 & 0.540 & 0.331 & 0.425 \\
\midrule
Qwen3-4B-Instruct & 0.138 & 0.497 & 0.482 & 0.285 \\
+SFT & 0.138 & 0.532 & \textbf{0.346} & 0.400 \\
+FinRankGRPO & \textbf{0.153} & \textbf{0.586} & 0.350 & \textbf{0.436} \\
\bottomrule
\end{tabular}
\caption{Results on the FF12-24 stock universe over the 2020--2025 out-of-sample period. Bold marks the strongest result among the three Qwen checkpoints. Null is the fraction of outputs that fail the complete-ranking parser; lower is better.}
\label{tab:stock_results}
\end{table}

Table~\ref{tab:stock_results} evaluates the operational transfer of the training pipeline to the individual-stock setting. The original Qwen3-4B-Instruct fails the complete-ranking parser on $99.4\%$ of its outputs and achieves a Spearman correlation of $-0.145$. SFT reduces the null rate to $5.2\%$, and FinRankGRPO further improves the annualized return to $0.153$ and the Sharpe ratio to $0.586$. FinRankGRPO improves over DeepSeek-Chat.

% Nevertheless, the Spearman correlation remains negative, and FinRankGRPO does not surpass the traditional financial baselines. One plausible explanation is that news matched only to the current trading date provides less firm-specific predictive information for individual stocks than for diversified ETFs, whose returns may respond more consistently to sector-level and macroeconomic developments. The model may therefore improve ranking quality relative to general-purpose LLMs without recovering a sufficiently accurate ordering of future stock returns. These results demonstrate that post training can substantially improve output validity and relative performance over the base LLM. Improving individual stock performance may require richer firm-specific signals and asset-specific training.

\section{Experiment Setup}
\label{appx:experiment_setup}

We present the experiment setup for FinRankGRPO across Base and Instruct models. All experiments were conducted using full-parameter training on 3 NVIDIA L20 GPUs with mixed precision (bfloat16) to optimize computational efficiency. Our torch version is 2.9.1, transformers is 4.57.1, and trl is 0.9.6. The model-scale study covers Qwen2.5 checkpoints at 0.5B, 1.5B, and 3B parameters; the main experiments use Qwen3-4B and Qwen3-4B-Instruct.

\textbf{SFT}. In this stage, the model is trained for 3 epochs using a learning rate of $10^{-4}$ about 1 hour. We adopt a cosine learning rate scheduler with 10\% warmup steps. Training is conducted with a per-device batch size of 1 and gradient accumulation over 4 steps. This configuration balances memory constraints with stable optimization, ensuring consistent and efficient fine-tuning across the dataset.

\textbf{FinRankGRPO}. Subsequently, the models were trained using the FinRankGRPO framework for 1 epoch, about 12 hours. This phase employed a learning rate of $3\times10^{-6}$, also with cosine decay and a 10\% linear warmup. Each device uses a batch size of 1 and gradient accumulation over 4 steps. During each training iteration, prompts with a maximum length of 1024 tokens and 4 responses were sampled per prompt for ranking based policy optimization. For training, the sampling parameters were temperature = 1.0, top-k = 50, and top-p = 1.0.

\section{Model Sizes and Release Dates}
\label{appx:model_information}

Table~\ref{tab:model_information} summarizes the models used for CoT supervision, commercial comparison, and open-weight post-training. Release dates follow the official announcements for GPT-4o-mini\footnote{\href{https://openai.com/index/gpt-4o-mini-advancing-cost-efficient-intelligence/}{Official GPT-4o-mini announcement.}}, Claude 3.5 Sonnet\footnote{\href{https://www.anthropic.com/news/3-5-models-and-computer-use}{Official Claude 3.5 Sonnet announcement.}}, DeepSeek-V3.2\footnote{\href{https://api-docs.deepseek.com/news/news251201/}{Official DeepSeek-V3.2 announcement.}}, Qwen2.5\footnote{\href{https://qwenlm.github.io/blog/qwen2.5/}{Official Qwen2.5 announcement.}}, and Qwen3\footnote{\href{https://qwenlm.github.io/blog/qwen3/}{Official Qwen3 announcement.}}. For commercial API models, we report ``--'' when the parameter count of the deployed endpoint is not publicly disclosed.

\begin{table}[t]
\centering
\footnotesize
\setlength{\tabcolsep}{4pt}
\renewcommand{\arraystretch}{1.10}
\caption{Parameter sizes and public release dates of the language models used in our experiments. ``--'' indicates that the parameter count of the deployed commercial API model was not publicly disclosed. Post-trained derivatives inherit the parameter size and release entry of their base checkpoint.}
\label{tab:model_information}
\begin{tabularx}{\columnwidth}{@{}Xcc@{}}
\toprule
\textbf{Model} & \textbf{Parameters} & \textbf{Public release date} \\
\midrule
GPT-4o-mini & -- & 2024-07-18 \\
Claude-3.5-Sonnet-20241022 & -- & 2024-10-22 \\
DeepSeek-Chat (V3.2) & -- & 2025-12-01 \\
Qwen2.5 Base/Instruct & 0.5B / 1.5B / 3B & 2024-09-19 \\
Qwen3-4B-Base & 4B & 2025-04-29 \\
Qwen3-4B-Instruct & 4B & 2025-04-29 \\
\bottomrule
\end{tabularx}
\end{table}

\section{Post-Release Cutoff Evaluation}
\label{sec:post_release_cutoff}

To reduce the risk that evaluation events appeared in foundation-model pretraining, we conduct an additional cutoff evaluation on weekly observations from June through December 2025. GPT-4o-mini has an official knowledge cutoff of October, 2023\footnote{\href{https://developers.openai.com/api/docs/models/gpt-4o-mini}{Official GPT-4o-mini model documentation.}} and was publicly released on July, 2024, while the fixed Qwen3-4B checkpoint was released on April 29, 2025. The entire evaluation interval therefore follows both models' disclosed cutoff or public release, and our SFT and FinRankGRPO data remain restricted to 2010--2019. We compare GPT-4o-mini with the Qwen3-4B base model and its two post-trained variants. Each portfolio begins with zero asset holdings, and the remaining evaluation protocol follows the main experiment.

\begin{table}[H]
\centering
\footnotesize
\setlength{\tabcolsep}{3.5pt}
\renewcommand{\arraystretch}{1.10}
\caption{Post-release cutoff results from June through December 2025. Each portfolio starts with zero asset holdings, and values are means over five runs. Bold marks the best value in each column.}
\label{tab:post_release_cutoff}
\begin{tabularx}{\columnwidth}{@{}Xrrrr@{}}
\toprule
\textbf{Method} & \textbf{Spear.} & \textbf{Kend.} & \textbf{Ann. Ret.} & \textbf{Sharpe} \\
\midrule
GPT-4o-mini & -0.021 & -0.025 & 0.152 & 0.953 \\
Qwen3-4B & 0.011 & 0.013 & -0.004 & -0.097 \\
\quad +SFT & 0.039 & 0.027 & 0.117 & 0.672 \\
\quad +FinRankGRPO & \textbf{0.047} & \textbf{0.039} & \textbf{0.171} & \textbf{1.159} \\
\bottomrule
\end{tabularx}
\end{table}

Table~\ref{tab:post_release_cutoff} shows that Qwen3-4B with FinRankGRPO performs best on all four reported metrics. It increases annualized return from $-0.004$ for the base model and $0.117$ for SFT to $0.171$, while increasing the Sharpe ratio from $-0.097$ and $0.672$ to $1.159$. It also reaches Spearman $0.047$ and Kendall's $\tau$ $0.039$, compared with $-0.021$ and $-0.025$ for GPT-4o-mini, and exceeds GPT-4o-mini in annualized return ($0.152$) and Sharpe ratio ($0.953$). This cutoff experiment therefore provides post-release evidence for both ranking accuracy and downstream portfolio utility in the evaluated interval.

\section{SFT Supervision Construction}
\label{appx:sft_supervision}

We examine whether the SFT gains depend on how teacher supervision is constructed. The full pipeline considers valid traces from GPT-4o-mini, Claude-3.5-Sonnet, and DeepSeek-Chat and, using only the 2010--2019 training labels, retains the trace whose final ranking has the highest Spearman correlation with the realized return ranking. The fixed-teacher variant uses the GPT-4o-mini trace whenever it is valid, and the random variant samples one valid commercial-model trace. The vote variant first forms a majority/Borda-style consensus ranking across the valid teacher rankings and then retains the original teacher trace whose ranking is closest to that consensus; it therefore avoids realized-return-based teacher selection while preserving a coherent teacher rationale. Finally, the answer-only variant starts from the same hindsight-best traces as full SFT but removes their reasoning and keeps only the final 15-ETF JSON ranking. It isolates the effect of chain-of-thought text rather than changing the target ranking.

Table~\ref{tab:sft_supervision_ablation} shows that answer-only supervision attains the strongest SFT-stage portfolio result ($0.108$ annualized return and $0.568$ Sharpe), while the full pipeline remains competitive ($0.107$ and $0.532$). Random teacher selection and consensus voting are weaker, with Sharpe ratios of $0.482$ and $0.514$, respectively. The SFT benefit therefore does not require retaining the teacher's written reasoning, but it remains sensitive to which target ranking is used.

\begin{table}[t]
\centering
\scriptsize
\renewcommand{\arraystretch}{1.08}
\caption{Ablation of SFT supervision construction for Qwen3-4B-Instruct. All entries are means over five runs. The full SFT row uses the main-experiment supervision pipeline.}
\label{tab:sft_supervision_ablation}
\resizebox{\columnwidth}{!}{%
\begin{tabular}{@{}lrrrr@{}}
\toprule
\textbf{SFT supervision} & \textbf{Spear.} & \textbf{Kend.} & \textbf{Ann. Ret.} & \textbf{Sharpe} \\
\midrule
Full SFT & 0.014 & 0.012 & 0.107 & 0.532 \\
Teacher voting & 0.011 & 0.009 & 0.102 & 0.514 \\
GPT-4o-mini only & 0.014 & \textbf{0.012} & 0.098 & 0.482 \\
Answer only & \textbf{0.015} & 0.012 & \textbf{0.108} & \textbf{0.568} \\
Random teacher response & 0.011 & 0.009 & 0.099 & 0.482 \\
\bottomrule
\end{tabular}
}
\end{table}

\section{Comparison with Alternative Ranking Methods}
\label{appx:other_methods}

Table~\ref{tab:other_methods} broadens the comparison beyond direct list generation. The two DeepSeek-V3.2 variants use the same API model, ETF universe, evaluation dates, and backtest. One-shot prompting, used here as a stronger prompting control than zero-shot inference, produces the complete ordering after one demonstration. We do not treat this single demonstration as a multi-shot sensitivity study. The point-wise variant instead asks the LLM to predict one numerical return per ETF, sorts those predictions into a complete ranking, and passes that ranking to the same entropy-pooling (EP) portfolio construction. It improves Spearman from $0.006$ to $0.013$ and Sharpe from $0.435$ to $0.524$, providing a controlled comparison of output formulation within the same commercial backbone.

The two non-LLM baselines use the same 2010--2019/2020--2025 temporal split and feed their predicted orders to the same EP and backtest. FinBERT--RankNet encodes the weekly news summary with FinBERT, maps the pooled representation to 15 asset scores, and learns all pairwise preferences with a RankNet loss; 2019 is held out for checkpoint selection. LambdaMART uses LightGBM's \texttt{lambdarank} objective with only lagged information available before the target week: asset identity, the previous return, its cross-sectional z-score and rank, its magnitude, and its sign. It uses 2019 to select the number of boosting rounds and is then refit on all 2010--2019 samples. The updated LambdaMART result reaches $0.015$ Spearman, $0.011$ Kendall, $0.069$ annualized return, and a $0.273$ Sharpe ratio, closely matching FinBERT--RankNet on rank correlation but producing weaker portfolio performance. FinBERT--RankNet attains the highest NDCG@3 ($0.575$), whereas FinRankGRPO achieves the strongest rank correlations ($0.018$ Spearman and $0.015$ Kendall), the best portfolio outcomes ($0.131$ annualized return and $0.689$ Sharpe), and the lowest maximum drawdown ($0.210$) among these methods.

\begin{table}[t]
\centering
\scriptsize
\renewcommand{\arraystretch}{1.08}
\caption{Comparison with prompting, point-wise prediction, pairwise classification, and learning-to-rank alternatives on ETF-15. All entries are means over five runs. Bold marks the best value in each column.}
\label{tab:other_methods}
\resizebox{\columnwidth}{!}{%
\begin{tabular}{@{}lrrrr@{}}
\toprule
\textbf{Method}
& \textbf{Spear.}
& \textbf{Kend.}
& \textbf{Ann. Ret.}
& \textbf{Sharpe} \\
\midrule
DeepSeek-V3.2 one-shot
& 0.006 & 0.004 & 0.090 & 0.435 \\
DeepSeek-V3.2 point-wise
& 0.013 & 0.011 & 0.105 & 0.524 \\
FinBERT pairwise
& 0.015 & 0.011 & 0.077 & 0.302 \\
LambdaMART
& 0.015 & 0.011 & 0.069 & 0.273 \\
FinRankGRPO
& \textbf{0.018}
& \textbf{0.015}
& \textbf{0.131}
& \textbf{0.689} \\
\bottomrule
\end{tabular}%
}
\end{table}

\section{Effect of Model Scale}
\label{appx:model_scale}

\begin{figure}[t]
    \centering
    \includegraphics[width=\columnwidth]{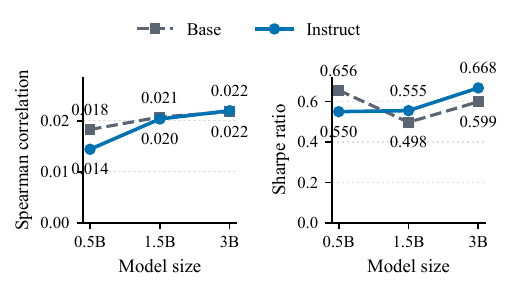}
    \caption{The 3B Qwen2.5 checkpoints perform best overall, although increasing model scale does not yield monotonic gains on every metric. All checkpoints use the same two-stage FinRankGRPO pipeline. Left: Spearman correlation. Right: Sharpe ratio.}
    \label{fig:model_comparison_base}
\end{figure}

Figure~\ref{fig:model_comparison_base} compares Qwen2.5 Base and Instruct checkpoints at 0.5B, 1.5B, and 3B parameters. For the Base checkpoints, Spearman correlation increases from $0.018$ at 0.5B to $0.022$ at 3B, whereas the Sharpe ratio decreases from $0.656$ to $0.498$ at 1.5B before recovering to $0.599$ at 3B. For the Instruct checkpoints, Spearman correlation increases from $0.014$ to $0.022$, and the Sharpe ratio increases from $0.550$ to $0.668$, with only a small gain at 1.5B ($0.555$). These results suggest that larger model generally improve rank correlation, but their effect on portfolio performance depends on the checkpoint family; notably, the 0.5B Base checkpoint achieves a higher Sharpe ratio than the 3B Base checkpoint.

\section{ Variability and Comparisons with Commercial APIs}
\label{appx:statistical_tests}

For every method in the main experiment, we report the mean and standard deviation. The 95\% confidence intervals use a two-sided Student $t$ interval with four degrees of freedom. 

\begin{table*}[t]
\centering
\scriptsize
\setlength{\tabcolsep}{4pt}
\renewcommand{\arraystretch}{1.16}
\begin{tabularx}{\linewidth}{@{}l*{4}{>{\centering\arraybackslash}X}@{}}
\toprule
\textbf{Model} & \textbf{Spearman} & \textbf{Kendall} & \textbf{NDCG@3} & \textbf{NDCG@5} \\
\midrule
DeepSeek-Chat & \shortstack{.0150 $\pm$ .0030\\(.0113, .0186)} & \shortstack{.0161 $\pm$ .0025\\(.0130, .0193)} & \shortstack{.5745 $\pm$ .0042\\(.5692, .5797)} & \shortstack{.6052 $\pm$ .0014\\(.6035, .6070)} \\
Claude & \shortstack{.0013 $\pm$ .0022\\(-.0014, .0040)} & \shortstack{.0045 $\pm$ .0018\\(.0023, .0066)} & \shortstack{.5700 $\pm$ .0029\\(.5663, .5736)} & \shortstack{.5975 $\pm$ .0020\\(.5950, .6000)} \\
GPT-4o-mini & \shortstack{.0141 $\pm$ .0035\\(.0098, .0185)} & \shortstack{.0131 $\pm$ .0035\\(.0088, .0174)} & \shortstack{.5704 $\pm$ .0035\\(.5661, .5748)} & \shortstack{.6039 $\pm$ .0009\\(.6028, .6050)} \\
\midrule
Qwen3-4B & \shortstack{-.0145 $\pm$ .2461\\(-.3201, .2911)} & \shortstack{-.0095 $\pm$ .1871\\(-.2419, .2228)} & \shortstack{.6011 $\pm$ .0826\\(.4986, .7036)} & \shortstack{.5818 $\pm$ .0889\\(.4713, .6922)} \\
+SFT & \shortstack{.0120 $\pm$ .0009\\(.0108, .0132)} & \shortstack{.0095 $\pm$ .0012\\(.0080, .0111)} & \shortstack{.5637 $\pm$ .0037\\(.5590, .5683)} & \shortstack{.5960 $\pm$ .0017\\(.5939, .5981)} \\
+FinRankGRPO & \shortstack{.0152 $\pm$ .0023\\(.0124, .0180)} & \shortstack{.0125 $\pm$ .0018\\(.0102, .0148)} & \shortstack{.5753 $\pm$ .0045\\(.5697, .5810)} & \shortstack{.6024 $\pm$ .0041\\(.5973, .6075)} \\
Qwen3-4B-Instruct & \shortstack{.0088 $\pm$ .0023\\(.0060, .0116)} & \shortstack{.0075 $\pm$ .0019\\(.0051, .0098)} & \shortstack{.5748 $\pm$ .0013\\(.5733, .5764)} & \shortstack{.6039 $\pm$ .0010\\(.6027, .6052)} \\
+SFT & \shortstack{.0144 $\pm$ .0030\\(.0107, .0181)} & \shortstack{.0118 $\pm$ .0021\\(.0091, .0144)} & \shortstack{.5719 $\pm$ .0033\\(.5678, .5759)} & \shortstack{.6009 $\pm$ .0028\\(.5974, .6043)} \\
+FinRankGRPO & \shortstack{.0182 $\pm$ .0025\\(.0151, .0213)} & \shortstack{.0150 $\pm$ .0016\\(.0130, .0169)} & \shortstack{.5708 $\pm$ .0039\\(.5659, .5757)} & \shortstack{.6022 $\pm$ .0015\\(.6004, .6041)} \\
\bottomrule
\end{tabularx}
\caption{Ranking metrics across five independent runs. Each cell reports mean $\pm$  standard deviation, with the 95\% $t$ confidence interval in parentheses.}
\label{tab:main_stats_ranking}
\end{table*}

\begin{table*}[t]
\centering
\scriptsize
\setlength{\tabcolsep}{4pt}
\renewcommand{\arraystretch}{1.16}
\begin{tabularx}{\linewidth}{@{}l*{4}{>{\centering\arraybackslash}X}@{}}
\toprule
\textbf{Model} & \textbf{Annualized Return} & \textbf{Sharpe} & \textbf{Maximum Drawdown} & \textbf{Calmar} \\
\midrule
DeepSeek-Chat & \shortstack{.1084 $\pm$ .0125\\(.0929, .1239)} & \shortstack{.5600 $\pm$ .0821\\(.4581, .6619)} & \shortstack{.2271 $\pm$ .0061\\(.2196, .2347)} & \shortstack{.4781 $\pm$ .0632\\(.3997, .5565)} \\
Claude & \shortstack{.1011 $\pm$ .0068\\(.0927, .1095)} & \shortstack{.3960 $\pm$ .0331\\(.3549, .4371)} & \shortstack{.2538 $\pm$ .0057\\(.2467, .2609)} & \shortstack{.3983 $\pm$ .0231\\(.3696, .4270)} \\
GPT-4o-mini & \shortstack{.0806 $\pm$ .0185\\(.0577, .1036)} & \shortstack{.3635 $\pm$ .1187\\(.2161, .5109)} & \shortstack{.2403 $\pm$ .0126\\(.2246, .2560)} & \shortstack{.3353 $\pm$ .0767\\(.2400, .4305)} \\
\midrule
Qwen3-4B & \shortstack{.0703 $\pm$ .0528\\(.0046, .1359)} & \shortstack{.2861 $\pm$ .7386\\(-.6309, 1.2032)} & \shortstack{.1615 $\pm$ .0592\\(.0880, .2350)} & \shortstack{.3888 $\pm$ .3139\\(-.0009, .7786)} \\
+SFT & \shortstack{.1095 $\pm$ .0065\\(.1014, .1176)} & \shortstack{.5585 $\pm$ .0407\\(.5079, .6091)} & \shortstack{.2358 $\pm$ .0168\\(.2149, .2567)} & \shortstack{.4667 $\pm$ .0501\\(.4044, .5289)} \\
+FinRankGRPO & \shortstack{.1112 $\pm$ .0306\\(.0731, .1492)} & \shortstack{.5739 $\pm$ .2106\\(.3124, .8354)} & \shortstack{.2197 $\pm$ .0219\\(.1925, .2469)} & \shortstack{.5120 $\pm$ .1633\\(.3092, .7147)} \\
Qwen3-4B-Instruct & \shortstack{.0942 $\pm$ .0031\\(.0903, .0980)} & \shortstack{.4306 $\pm$ .0192\\(.4068, .4545)} & \shortstack{.2600 $\pm$ .0015\\(.2581, .2619)} & \shortstack{.3621 $\pm$ .0112\\(.3482, .3760)} \\
+SFT & \shortstack{.1065 $\pm$ .0205\\(.0811, .1320)} & \shortstack{.5323 $\pm$ .1296\\(.3713, .6932)} & \shortstack{.2217 $\pm$ .0105\\(.2087, .2347)} & \shortstack{.4839 $\pm$ .1107\\(.3464, .6214)} \\
+FinRankGRPO & \shortstack{.1309 $\pm$ .0058\\(.1236, .1381)} & \shortstack{.6885 $\pm$ .0419\\(.6365, .7405)} & \shortstack{.2099 $\pm$ .0114\\(.1957, .2241)} & \shortstack{.6249 $\pm$ .0425\\(.5721, .6777)} \\
\bottomrule
\end{tabularx}
\caption{Portfolio metrics across five independent runs. Each cell reports mean $\pm$ standard deviation, with the 95\% $t$ confidence interval in parentheses. Maximum drawdown is reported as an absolute magnitude.}
\label{tab:main_stats_portfolio}
\end{table*}

Qwen3-4B-Instruct with FinRankGRPO has a Spearman mean of $0.0182$ (95\% CI $[0.0151, 0.0213]$) and a Sharpe mean of $0.6885$ (95\% CI $[0.6365, 0.7405]$).

\begin{table}[t]
\centering
\scriptsize
\resizebox{\columnwidth}{!}{%
\begin{tabular}{@{}lrrrrrr@{}}
\toprule
\textbf{Comparator}
& $\Delta$\textbf{ Spear.} & $p$
& $\Delta$\textbf{ Ann. Ret.} & $p$
& $\Delta$\textbf{ Sharpe} & $p$ \\
\midrule
Standard GRPO
& 0.0047 & 0.13
& 0.0396 & 0.00
& 0.2526 & 0.01 \\
DeepSeek-Chat
& 0.0032 & 0.04
& 0.0225 & 0.01
& 0.1285 & .03 \\
Claude
& 0.0169 & 0.00
& 0.0298 & 0.01
& 0.2925 & 0.00 \\
GPT-4o-mini
& 0.0041 & 0.16
& 0.0503 & 0.00
& 0.3250 & 0.00\\
\bottomrule
\end{tabular}%
}
\caption{Two-sided paired $t$-tests comparing Qwen3-4B-Instruct + FinRankGRPO with standard GRPO and commercial APIs across five runs.}
\label{tab:significance_tests}
\end{table}

We compare the best FinRankGRPO configuration with standard GRPO and each commercial API using two-sided paired $t$-tests over the five runs. FinRankGRPO has significantly higher annualized return and Sharpe ratio in all four comparisons.
Its Spearman advantage is significant against DeepSeek-Chat and Claude, but not against standard GRPO or GPT-4o-mini.

\section{Validity of the Learned Ranking Signal}
\label{appx:answer_perturbation}

We test whether the downstream results depend on the ordering learned by FinRankGRPO rather than merely on passing a complete list of assets to Entropy Pooling. Starting from Qwen3-4B-Instruct with FinRankGRPO, we replace its predicted order with either a seed-42 random permutation, the reverse of the predicted order, or the realized ground-truth order. Table~\ref{tab:answer_perturbation} compares these controls with the FinRankGRPO result from the main experiment.

\begin{table}[t]
\centering
\footnotesize
\setlength{\tabcolsep}{2.8pt}
\renewcommand{\arraystretch}{1.12}
\begin{tabularx}{\columnwidth}{@{}Xrrrr@{}}
\toprule
\textbf{Variant} & \textbf{Spear.} & \textbf{Kend.} & \textbf{Return} & \textbf{Sharpe} \\
\midrule
FinRankGRPO & \textbf{0.018} & \textbf{0.015} & \textbf{0.131} & \textbf{0.689} \\
Random ranking & 0.005 & 0.004 & 0.052 & 0.070 \\
Reversed FinRankGRPO ranking & -0.018 & -0.015 & 0.052 & 0.119 \\
Ground-truth ranking + EP & 1.000 & 1.000 & 1.049 & 7.653 \\
\bottomrule
\end{tabularx}
\caption{Effect of the ranking supplied to the same EP and portfolio pipeline. The first three learned or perturbed variants are means over five runs. Ground-truth ranking + EP uses the realized return order and is a non-deployable oracle. Return denotes annualized return. Bold marks the best deployable value in each column.}
\label{tab:answer_perturbation}
\end{table}

Both perturbations sharply reduce ranking accuracy and portfolio performance. Relative to FinRankGRPO, randomization reduces Spearman correlation from $0.018$ to $0.005$, Kendall's $\tau$ from $0.015$ to $0.004$, annualized return from $0.131$ to $0.052$, and the Sharpe ratio from $0.689$ to $0.070$. Reversing the learned order changes the rank correlations from positive to negative and yields an annualized return of $0.052$ and a Sharpe ratio of $0.119$. Conversely, the ground-truth ranking + EP oracle reaches $1.049$ annualized return and $7.653$ Sharpe. Holding EP fixed while moving from random or reversed rankings through the learned ranking to the oracle therefore isolates ranking quality as a material driver of downstream performance.

\section{Qualitative Success and Failure Cases}
\label{appx:case_analysis}

We select the highest- and lowest-scoring valid FinRankGRPO outputs after recomputing Spearman against the canonical realized ranking. Table~\ref{tab:case_analysis} shows that the successful case captures a broad risk-on regime and correctly orders both the leading equity sectors and defensive tail. The failure case illustrates a recurring limitation: salient narrative evidence can be directionally misleading over a one-week horizon. Negative energy and fossil-fuel headlines led the model to rank XLE and GSG last, although they became the two strongest realized assets. The examples therefore complement the aggregate metrics by showing both semantic sensitivity and event-to-return uncertainty.

\begin{table}[t]
\centering
\footnotesize
\setlength{\tabcolsep}{4pt}
\renewcommand{\arraystretch}{1.10}
\caption{High- and low-scoring FinRankGRPO examples selected by weekly Spearman correlation. Rankings are abbreviated to the top and bottom three assets; the score is computed on the complete 15-ETF order.}
\label{tab:case_analysis}
\begin{tabularx}{\columnwidth}{@{}l X@{}}
\toprule
\multicolumn{2}{@{}l}{\textbf{Successful case: 2024-08-12 \quad Spear. 0.864}} \\
\midrule
Predicted & XLK, XLY, IVV / GLD, XLU, GSG \\
Realized & XLK, XLY, IVV / GLD, XLE, GSG \\
Interpretation & The response recognized a broad risk-on rally and correctly placed technology, discretionary, and large-cap equities at the top. Its main tail error was ranking utilities below energy. \\
\midrule
\multicolumn{2}{@{}l}{\textbf{Failure case: 2021-08-23 \quad Spear. $-0.807$}} \\
\midrule
Predicted & GLD, XLK, XLV / XLB, XLE, GSG \\
Realized & GSG, XLE, IJH / XLV, XLP, XLU \\
Interpretation & The response extrapolated negative fossil-fuel headlines and a risk-off narrative, placing commodities and energy last. Realized returns reversed this narrative, with GSG and XLE leading the week. \\
\bottomrule
\end{tabularx}
\end{table}

\section{Entropy Pooling Example}
\label{appx:ep_example}

We use a prediction from the 2024-08-12 record in our dataset to illustrate the EP construction. The three highest-ranked assets are $\text{XLK} \succ \text{XLY} \succ \text{IVV}$. Restricting the illustration to these three assets gives the views $\mathbb{E}[R_{XLK}] \geq \mathbb{E}[R_{XLY}]$ and $\mathbb{E}[R_{XLY}] \geq \mathbb{E}[R_{IVV}]$. The corresponding pick matrix is:
\begin{equation}
\underbrace{\begin{bmatrix} 
1 & -1 & 0 \\ 
0 & 1 & -1 
\end{bmatrix}}_{\mathbf{P}} 
\begin{bmatrix} 
\mathbb{E}[R_{XLK}] \\ \mathbb{E}[R_{XLY}] \\ \mathbb{E}[R_{IVV}]
\end{bmatrix} 
\geq 
\underbrace{\begin{bmatrix} 
0 \\ 0 
\end{bmatrix}}_{\mathbf{v}}.
\end{equation}

\section{Portfolio Optimization}
\label{appx:details_portfolio_optimization}

% In this study, we implement two distinct portfolio optimization approaches. One based on the views generated by LLMs and the others use traditional mean-variance optimization as a baseline. The objective is to compare the performance of LLM-based portfolios against a benchmark portfolio optimized using conventional technique.

We compute the posterior expected-return vector and covariance matrix from the entropy-pooled scenario distribution and use them in a fully invested, long-only maximum-Sharpe optimization. Portfolio risk is measured by the standard deviation of returns. The annual risk-free rate is fixed at \(2.5\%\) and converted to the daily rate \(0.025/252\) to match the return frequency. We impose nonnegative weights and the full-investment constraint, without an explicit turnover penalty or additional concentration constraint. After optimization, small negative numerical residuals are clipped to zero, and the resulting weights are normalized to sum to one.
\section{ Detailed  Statistics}
\label{appx:detailed_statistics}

\begin{enumerate}
    \item \textbf{Annualized Return (Ann. Ret.)}: The geometric average amount of money earned by an investment each year over a given time period.
    % \[ R_{Ann} = \left( \prod_{t=1}^{T} (1 + r_t) \right)^{252/T} - 1 \]
    
    % \item \textbf{Annualized Volatility (Ann. Vol.)}: A statistical measure of the dispersion of returns, representing risk.
    % % \[ \sigma_{Ann} = \sqrt{\frac{252}{T-1} \sum_{t=1}^{T} (r_t - \bar{r})^2} \]
    
    \item \textbf{Sharpe Ratio}: Measures risk-adjusted performance. A higher Sharpe ratio indicates better returns for the same level of risk.
    % \[ \text{Sharpe} = \frac{R_{Ann} - R_f}{\sigma_{Ann}} \]
    % where $R_f$ is the risk-free rate (set to be 2.5\%).
    
    \item \textbf{Maximum Drawdown (Max DD)}: The maximum observed loss from a peak to a trough of a portfolio, before a new peak is attained.
    % \[ \text{Max DD} = \min_{t} \left( \frac{V_t}{\max_{\tau \in [0,t]} V_\tau} - 1 \right) \]
    
    \item \textbf{Calmar Ratio}: A measure of risk-adjusted return based on the Maximum Drawdown.
    % \[ \text{Calmar} = \frac{R_{Ann}}{|\text{Max DD}|} \]
\end{enumerate}

% For the backtest evaluation, we perform weekly portfolio rebalancing based on the generated weights. The environment is configured to simulate realistic trading conditions. We impose a transaction fee of 5 basis points (0.05\%) per turnover to account for commissions and bid-ask spreads. Trades are assumed to be executed at the weekly closing price. Given the high liquidity of the top-tier ETFs, we assume zero slippage for the execution. Furthermore, management fees are excluded from the calculation to strictly isolate the alpha generation capability of the active ranking strategy.

For the backtest evaluation, portfolios are rebalanced weekly under long-only constraints based on the generated weights. We use a simplified trading setting in which trades are executed at the weekly closing price and transaction costs of 5 basis points (0.05\%) per unit of portfolio turnover are deducted. Given the high liquidity of the selected ETFs, we assume zero slippage and exclude management fees. Model outputs that do not contain a complete valid ranking are excluded from the computation of ranking metrics. On these weeks, no rebalancing trade is executed, and the previous portfolio holdings are carried forward.
\section{Recession Regime}

Table \ref{tab:single_column_comparison} shows the model answer for recession Regime
\begin{table}[htbp]
\centering
\small
\begin{tabular}{p{\dimexpr\linewidth-2\tabcolsep\relax}}
\toprule
% \textbf{Market Regime} \\
% \midrule
\textbf{Market Regime: Recession} \\
\textit{News: China's Premier Li Says Coronavirus Outbreak Control Remains at Crucial Stage; JD.com Trades Higher On Q4 Results, Announces CFO Succession; Fed Cuts Rates By 0.5\%; Daily Markets: Investors Shrug Off Fed's Emergency Rate Cut; Apple May Soon Stop Using Intel Chips in Macs $\dots$} \\
\textit{Ground-truth Ranking:} "[GSG, \textbf{GLD}, XLP, XLV, XLK, IVV, XLF, XLB, XLI, XLY, XLE, IJR, IJH, XLU, IYR]" \\
\midrule
\textbf{Commercial (DeepSeek-Chat)} \\
\textit{Ranking:} "[\textbf{GLD}, XLV, XLU, XLP, IVV, XLK, IYR, IJH, XLB, XLF, XLY, XLI, GSG, XLE, IJR]" \\
% \textbf{Commercial (GPT-4o-mini)} \\
% \textit{Ranking:} "[\textbf{GLD}, XLU, XLP, IVV, XLK, XLV, IYR, XLB, GSG, XLY, XLI, XLE, XLF, IJH, IJR]" \\
\midrule
\textbf{Qwen3-4B-Instruct (Base Model)} \\
\textit{Ranking:} "[XLK, XLV, \textbf{GLD}, XLP, XLU, XLF, XLY, XLE, XLI, XLB, IVV, IJH, IJR, GSG, IYR]" \\
\midrule
\textbf{Qwen3-4B-Instruct (SFT)} \\
\textit{Ranking:} "[XLK, \textbf{GLD}, XLV, XLU, XLP, IVV, XLI, XLB, IJH, XLY, IYR, XLF, IJR, XLE, GSG]" \\
\midrule
\textbf{Qwen3-4B-Instruct (FinRankGRPO Ours)} \\
\textit{Ranking:} "[XLK, \textbf{GLD}, XLV, XLP, XLU, IVV, XLI, IJH, XLB, IJR, XLF, XLE, IYR, GSG, XLY]" \\
\bottomrule
\end{tabular}
\caption{Models' predicted ranking lists for the week of 2020-03-09 and the ground-truth ranking. GLD is bolded for comparison.}
\label{tab:single_column_comparison}

\end{table}

\section{Sentiment Reversal Example}

Table~\ref{tab:sentiment_reversal_example} presents an example of the sentiment-reversal procedure.
\label{appx:sentiment}

\begin{table}[t]
\centering
\scriptsize
\renewcommand{\arraystretch}{0.94}
\begin{tabularx}{\columnwidth}{@{}X@{}}
\toprule
\textbf{Original News Summary} \\
\midrule
Based on the provided news titles from December 30, 2019, the dominant themes are mixed corporate earnings, sector-specific developments, and geopolitical tensions, against a backdrop of modest market \textbf{weakness} indicated by a \textbf{declining} Dow. The impact across the 15 specified ETFs will be varied and largely indirect.

\textbf{Sector ETFs:} The news is most directly impactful for \textbf{Technology (XLK)}, with \textbf{positive} updates from Netflix, NVIDIA, and Tesla's Gigafactory, though tempered by notable \textbf{outflows} in semiconductor-related funds. \textbf{Consumer Discretionary (XLY)} faces crosscurrents: \textbf{positive} Chinese EV news (Nio) contrasts with retail \textbf{distress} (Gap, Signet Jewelers, SPDR S\&P Retail ETF \textbf{outflows}). \textbf{Consumer Staples (XLP)} shows clear \textbf{negative} pressure, highlighted by repeated headlines of significant fund \textbf{outflows}. \textbf{Energy (XLE)} may see volatility due to headlines on political \textbf{unrest} impacting oil and a specific stake \textbf{sale} (TC Energy), while \textbf{Health Care (XLV)} gets a mild \textbf{boost} from FDA \textbf{approval} news (Myriad). Other sectors like Industrials, Financials, and Utilities lack direct catalysts.

\textbf{Market Cap ETFs:} The blend of large-cap tech strength and broader retail/industrial \textbf{sluggishness} suggests a neutral to slightly \textbf{negative} impact on \textbf{Large Cap (IVV)}. The \textbf{underperformance} of specific retailers and transports (Union Pacific) points to potential \textbf{headwinds} for \textbf{Small Cap (IJR)} and \textbf{Mid Cap (IJH)}, which are more exposed to domestic consumer and industrial cycles.

\textbf{Commodity ETFs:} Geopolitical tension from the Iran headline provides a classic short-term \textbf{bid} for \textbf{Gold (GLD)} as a \textbf{safe-haven} asset. \textbf{Broad Commodities (GSG)} may see \textbf{muted} effects, balancing potential oil volatility with other commodity news like China's pork reserve release.

Overall, the news flow supports a \textbf{defensive} tilt, favoring sectors with idiosyncratic \textbf{positive} news (Tech, Health Care) and \textbf{safe-haven} assets (Gold), while highlighting \textbf{weakness} in consumer-facing segments. \\
\midrule
\textbf{Sentiment-Reversed News Summary} \\
\midrule

Based on the provided news titles from December 30, 2019, the dominant themes are mixed corporate earnings, sector-specific developments, and geopolitical tensions, against a backdrop of modest market \textbf{strength} indicated by a \textbf{rising} Dow. The impact across the 15 specified ETFs will be varied and largely indirect.

\textbf{Sector ETFs:} The news is most directly impactful for \textbf{Technology (XLK)}, with \textbf{negative} updates from Netflix, NVIDIA, and Tesla's Gigafactory, though tempered by notable \textbf{inflows} in semiconductor-related funds. \textbf{Consumer Discretionary (XLY)} faces crosscurrents: \textbf{negative} Chinese EV news (Nio) contrasts with retail \textbf{resilience} (Gap, Signet Jewelers, SPDR S\&P Retail ETF \textbf{inflows}). \textbf{Consumer Staples (XLP)} shows clear \textbf{positive} pressure, highlighted by repeated headlines of significant fund \textbf{inflows}. \textbf{Energy (XLE)} may see \textbf{stability} due to headlines on political \textbf{calm} impacting oil and a specific stake \textbf{purchase} (TC Energy), while \textbf{Health Care (XLV)} gets a mild \textbf{setback} from FDA \textbf{rejection} news (Myriad). Other sectors like Industrials, Financials, and Utilities lack direct catalysts.

\textbf{Market Cap ETFs:} The blend of large-cap tech \textbf{weakness} and broader retail/industrial \textbf{strength} suggests a neutral to slightly \textbf{positive} impact on \textbf{Large Cap (IVV)}. The \textbf{outperformance} of specific retailers and transports (Union Pacific) points to potential \textbf{tailwinds} for \textbf{Small Cap (IJR)} and \textbf{Mid Cap (IJH)}, which are more exposed to domestic consumer and industrial cycles.

\textbf{Commodity ETFs:} Geopolitical \textbf{calm} from the Iran headline provides a classic short-term \textbf{decline} for \textbf{Gold (GLD)} as a \textbf{risk-on} asset. \textbf{Broad Commodities (GSG)} may see \textbf{pronounced} effects, balancing potential oil \textbf{stability} with other commodity news like China's pork reserve \textbf{withdrawal}.

Overall, the news flow supports an \textbf{aggressive} tilt, favoring sectors with idiosyncratic \textbf{negative} news (Tech, Health Care) and \textbf{risk-on} assets (Gold), while highlighting \textbf{strength} in consumer-facing segments. \\
\bottomrule
\end{tabularx}

\caption{Example of the sentiment reversal procedure for the trading date 2020-01-06. Bold text highlights the transformed sentiment expressions. }
\label{tab:sentiment_reversal_example}
\end{table}

\section{Portfolio Net Value Comparison}

\begin{figure*}[t]
  \centering
  \includegraphics[width=.9\textwidth]{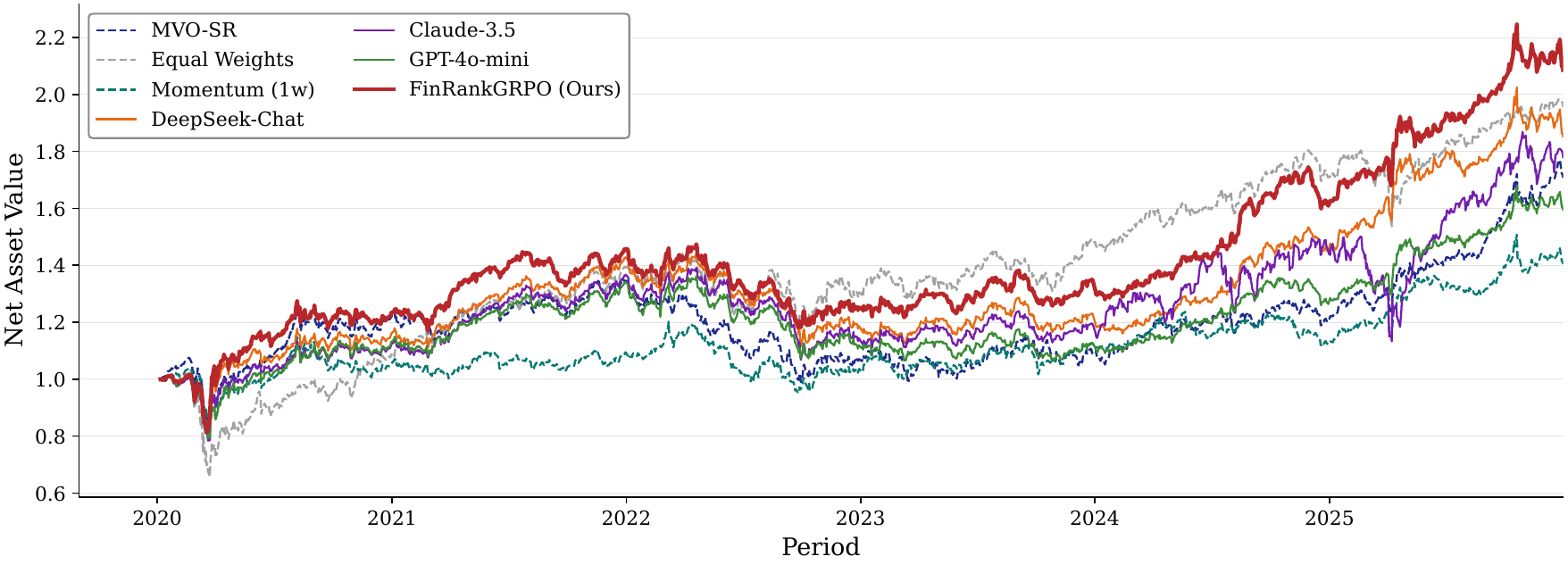}
  \caption{FinRankGRPO attains the highest terminal net value in the studied 2020--2025 setting. Curves compare traditional baselines and LLM ranking with EP methods. The year labels on the x-axis mark the beginning of each year (i.e., January 1).}
  \label{fig:portfolio_performance}
\end{figure*}

Figure~\ref{fig:portfolio_performance} compares the cumulative net asset values of the evaluated strategies.

\label{appx:portfolio_nav}

\end{document}